\documentclass[onluarkali,ingilizce,lisans,karton,ehb,tarihli,onsozacik,ozetacik,appendix]{itutez}

\usepackage{amsmath}

\DeclareMathOperator*{\argmax}{arg\,max}

\usepackage{lipsum}
\usepackage{graphicx}
\usepackage{afterpage}
\usepackage{lscape}
\usepackage{float}
\usepackage{mathtools}
\usepackage{listings}
\usepackage[final]{pdfpages}
\usepackage{hyperref}
\usepackage{url}
\usepackage{listings}
\usepackage{amssymb}

\yazar{Temp}{Author}{150230423}
\authorSecond{Kutay}{BÖLAT}{(040140012)}
\authorFirst{}{}{}
\authorThird{}{}{}

\unvan{}
 
\anabilimdali{}{Department of Control and Automation Engineering}
\programi{}{Kontrol ve Otomasyon Mühendisliği Bölümü}

\tarih{TEMMUZ, 2020}{JULY, 2020}
\tarihKucuk{TEMMUZ, 2020}{JULY, 2020}

\tezyoneticisi{Assoc. Prof. Dr. Tufan KUMBASAR}{Doç. Dr. Tufan KUMBASAR}  
\esdanismani{}{}

\baslik{CHANNEL STATE INFORMATION BASED} {LOCALIZATION WITH DEEP LEARNING}{}
\title{DERİN ÖĞRENME İLE KANAL DURUM BİLGİSİ} {TEMELLİ LOKALİZASYON}{}

\usepackage{color}
\usepackage{times}
\usepackage{amssymb,amsmath,mathptmx,amsbsy,bm}
\usepackage{caption}            
\usepackage{graphics}
\usepackage{wrapfig}
\usepackage{epsfig}
\usepackage{enumerate}
\usepackage{rotating}
\usepackage{multirow}
\usepackage{subfigure}
\usepackage{colortbl}
\usepackage{pstricks}
\usepackage{pst-plot}
\usepackage{cite}%
\usepackage{latexsym}           %
\usepackage{rotating}
\usepackage{fixltx2e} 

\def\be{\begin{equation}} %
\def\ee{\end{equation}}%
\def\beq{\begin{eqnarray}}%
\def\eeq{\end{eqnarray}}%
\def\bse{\begin{subequations}}%
\def\ese{\end{subequations}}%
\def\[{\left[}
\def\]{\right]}
\def\({\left(}
\def\){\right)}

\ithaf{}

\kisaltmalistesi{\hspace{-3mm}
\begin{tabular}{p{2cm}l}

{\bf AWGN} & {\bf:} Additive White Gaussian Noise\\
{\bf CL} & {\bf:} Convolutional Layer\\
{\bf CSI} & {\bf:} Channel State Information\\
{\bf FCL} & {\bf:} Fully Connected Layer\\
{\bf GPS} & {\bf:} Global Positioning System\\
{\bf IFFT} & {\bf:} Inverse Fast Fourier Transform\\
{\bf IP} & {\bf:} Internet Protocol\\
{\bf LOS} & {\bf:} Line of Sight\\
{\bf MAC} & {\bf:} Media Access Control\\
{\bf NLOS} & {\bf:} Non Line of Sight\\
{\bf OFDM} & {\bf:} Orthogonal Frequency Division Multiplexing\\
{\bf OpenWrt} & {\bf:} OPEN Wireless Router\\
{\bf PC} & {\bf:} Personal Computer\\
{\bf PDF} & {\bf:} Probability Density Function\\
{\bf ReLU} & {\bf:} Rectified Linear Unit\\
{\bf RMSProp} & {\bf:} Root Mean Squared Propagation\\
{\bf SGDM} & {\bf:} Stochastic Gradient Descent with Momentum\\
{\bf SSH} & {\bf:} Secure Shell\\
{\bf UAV} & {\bf:} Unmanned Aerial Vehicle\\

\end{tabular}

}
\sembollistesi{\hspace{-3mm}
\begin{tabular}{p{2cm}l}

{$\sigma$ } & {\bf:} Standart deviation\\
{$\alpha$ } & {\bf:} Path loss\\
{$\phi$ }  & {\bf:} Phase shift\\
{$h$} & {\bf:} Channel coefficient\\
{$c$} & {\bf:} Speed of light\\
{$\Omega$} & {\bf:} Complex CSI matrix\\
{$\bar\Omega$} & {\bf:} Normalized CSI matrix\\
{$\mathbf{W}$} & {\bf:} Weight matrix\\
{$\mathbf{b}$} & {\bf:} Bias vector\\
{$\pi$} & {\bf:} Category\\
{$J$} & {\bf:} Loss value\\
{$\hat\Omega$} & {\bf:} Unity power CSI matrix\\
{$o$} & {\bf:} Convolutional layer output size\\
{$i$} & {\bf:} Convolutional layer input size\\
{$f$} & {\bf:} Convolutional layer filter size\\
{$s$} & {\bf:} Convolutional layer stride size\\
{$\gamma$} & {\bf:} Leaky ReLU scaling parameter\\
{$\beta_1$} & {\bf:} Gradient decay rate\\
{$\beta_2$} & {\bf:} Squared gradient decay rate\\
{$\theta$} & {\bf:} Learnable parameter\\
{$\eta$} & {\bf:} Learning rate\\
{$\epsilon$} & {\bf:} Arbitrarily small number\\
{$m$} & {\bf:} First moment estimation of gradients\\
{$v$} & {\bf:} Second moment estimation of gradients\\

\end{tabular}

}
\onsoz{I would like to thank my advisor Assoc. Prof. Tufan KUMBASAR for his support and guidance throughout the graduation project period. I am grateful that Artificial Intelligence and Intelligent Systems (AI2S) Laboratory opened its doors and provided a healthy study environment for this project. This research is supported by the project (118E807) of Scientific and
Technological Research Council of Turkey.

I also thank my beloved family and dearest friends for their endless support and undeniable trust during the semester.
}             
\summary{Lokalizasyon, robotik ve telsiz haberleşme gibi çeşitli alanlarda en önemli sorunlardan biridir. Örneğin, İnsansız Hava Araçları (İHA), uygun bir kontrol stratejisi için hassa bir konum bilgisi gerektirir. Dış mekan uygulamaları için entegre GPS üniteleri ile bu problem çok verimli bir şekilde ele alınmaktadır. Ancak, iç mekan uygulamaları GPS sinyallerinin eksikliği nedeniyle özel bir uygulama gerektirir. Bu uygulama çoğunlukla görsel sistemler ve özel veri işleme yazılımları ile gerçekleştirilir.

İHA'lar gibi mobil robotların başka bir yönü de mobil robot ile kişisel bilgisayar gibi bir hesaplama birimi arasında sürekli bir kablosuz haberleşme olmasıdır. Bu haberleşme esas olarak doğrudan telemetri bilgisinin eldesi veya kontrol eylemlerinin hesaplanması için yapılabilir. Bu iletim için sorumlu entegre birimler, IEEE 802.11ac gibi belirli standartlar altında çalışmak üzere tasarlanmış ticari kablosuz haberleşme yonga kümeleridir.

Herhangi bir haberleşme sisteminin temel amacı, güç, zaman ve bant genişliği gibi sınırlı miktarda kaynakla bilgi iletmektir. Bununla birlikte, iletilen bilgi bitleri, kablosuz kanal etkileri nedeniyle  bozulabilir ve hatalı bir şekilde çözülebilir. Alıcı birim, çeşitli matematiksel tekniklerle kanalın bu çeşitli etkilerinden kurtulmaktan sorumludur. Bu teknikler esas olarak kanalın kendisini kompanze etmek için mevcut kanalın Kanal Durumu Bilgisini (channel state information, CSI) gerektirir. Bu kompanzasyondan sonra yonga setinin CSI ile bir ilişiği kalmaz. Ancak, kablosuz kanal çevreye göre değişen bir etmendir. Bu nedenle, hem vericinin hem de alıcının anlık konumları CSI üzerinde doğrudan bir etkiye sahiptir.

CSI, çevre hakkında bu kadar zengin bilgiler içermesine rağmen, kullanıcıya yalnızca işlenmiş bilgi veri bitlerini sağlamak için üretildiğinden, bu verilerin erişilebilirliği ticari kablosuz yonga setleri tarafından sınırlandırılmıştır. Bununla birlikte, IEEE 802.11n standardıyla, belirli yonga setleri CSI'ya erişim sağlamaktadır. Bu nedenle, CSI verileri işlenebilir ve lokalizasyon sistemleriyle bütünleştirilebilir hale gelmiştir.

Derin Öğrenme, yüksek boyutlu verileri işlemek için yaygın ve güçlü bir yöntemdir. Gücünü, verilerdeki son derece doğrusal olmayan ilişkileri yakalama ve bunu insani bir manipülasyon olmadan yapma yeteneğine borçludur. Bu nedenle, kablosuz kanal özelliklerini kullanarak yapılacak bir lokalizasyon uyulaması için Derin Öğrenme'nin kullanılması uygun görünmektedir.

Bu projede, lokalizasyon problemi için bir test ortamı oluşturulmuştur. Uygun yonga setlerine sahip iki yönlendirici, verici ve alıcı olarak atanmıştır. Ardından bu üniteler CSI verisini toplamalerı için operasyonel hale getirilmiştir. Son olarak, bu veriler çeşitli Derin Öğrenme modelleri ile işlenmiştir.

}            
\ozet{Localization is one of the most important problems in various fields such as robotics and wireless communications. For instance, Unmanned Aerial Vehicles (UAVs) require the information of the position precisely for an adequate control strategy. This problem is handled very efficiently with integrated GPS units for outdoor applications. However, indoor applications require special treatment due to the unavailability of GPS signals. This treatment is mostly conducted via visual systems and specialized processing software.

Another aspect of mobile robots such as UAVs is that there is constant wireless communication between the mobile robot and a computational unit such as a personal computer. This communication is mainly done for obtaining telemetry information or computation of control actions directly. The responsible integrated units for this transmission are commercial wireless communication chipsets which are designed to operate under certain standards such as IEEE 802.11ac.

The main purpose of any communication system is to transmit information with a limited amount of resources such as power, time, and bandwidth. However, the transmitted information bits can be corrupted and decoded erroneously due to the wireless channel effects. A receiver unit is responsible for getting rid of these diverse effects of the channel with various mathematical techniques. These techniques mainly require the Channel State Information (CSI) of the current channel to compensate the channel itself. After the compensation, the chipset has nothing to do with CSI. However, this wireless channel is a notion that alters with the environment. Therefore, the locations of both the transmitter and receiver have a direct impact on CSI.

Even though CSI contains such rich information about the environment, the accessibility of these data is blocked by the commercial wireless chipsets since they are manufactured to provide only the processed information data bits to the user. However, with the IEEE 802.11n standardization, certain chipsets provide access to CSI. Therefore, CSI data became processible and integrable to localization schemes.

Deep Learning is a common and powerful method to process high dimensional data. It owes its power to its ability to capture highly nonlinear relations among the data and do this without human tailoring. Therefore, using Deep Learning to capture the wireless channel characteristics for localization seems appropriate.

In this project, a test environment was constructed for the localization task. Two routers with proper chipsets were assigned as transmitter and receiver. They were operationalized for the CSI data collection. Lastly, these data were processed with various Deep Learning models.}

\begin{document}

\chapter{INTRODUCTION}

In today's world, the importance of localization gets bigger and bigger. This is due to the real-time applications in various fields that require, precise or imprecise, position information. The designers and engineers of these fields seek better and more flexible solutions for their localization problems day by day.

One of the most efficient localization techniques is the Global Positioning System (GPS) and it is integrated almost every mobile unit nowadays. However, the main drawback of this system is that it is unavailable indoor, due to the fact that GPS mainly utilizes a satellite network information stream and the electromagnetic waves carry this stream attenuated drastically while penetrating concrete walls. Thus, a different approach is followed in indoor applications. This approach generally follows the process of first identifying the global frame of the environment and then estimating the position of the mobile unit in this global frame. The systems of this type are generally used for the area of mobile robotics for efficient navigation and control. A great example of this change of systems between indoor and outdoor is unmanned aerial vehicles (UAVs). These mobile robots are known for their vast outdoor exploration area thanks to their GPS integration. However, visual localization systems with multiple cameras are utilized in indoor applications due to the lack of GPS signals.

One of the other aspects of UAVs, or mobile robots in general, is the fact that they constantly communicate with a ground station such as a personal computer (PC). This communication is done generally to send telemetry data to the user. Also, there exist mobile units that leave the control algorithms to the ground station and take the control inputs wirelessly. During these processes, the units follow a certain communication standard dictated by the integrated communication chipsets. These standards available today provide only a limited amount of information about the processes done in the physical layer of the communication systems. This information mostly consists of channel center frequency, bandwidth, and received signal strength indicator (RSSI). Access to the deeper levels of raw communication signals, require more sophisticated hardware such as software-defined radios.

A wireless communication channel can be thought of as a mathematical model of the random behavior of the communication signals in the environment. One of the most important parts of the physical layer of a communication system is the estimation of this channel. This estimation is crucial to compensate unwanted effects of the distorted wireless channel. In other words, the communication performance is heavily dependent on how successful the channel estimation is done. In contemporary communication systems, where \emph{orthogonal frequency division multiplexing} (OFDM) is heavily used, this channel estimation is done by acquiring the \emph{channel state information} (CSI).

The wireless channel is heavily affected by the environment due to electromagnetic phenomenons such as scattering. Therefore, CSI can be thought of as a probabilistic indicator of the location of the mobile unit with respect to the ground station in a stationary environment. However, most of the communication chipsets in the market remove the CSI data after decoding the information bits. Luckily, with the IEEE 802.11n standardization, certain chipset providers included CSI acquisition for the end-user. This advancement has opened the doors for more sophisticated localization techniques with CSI data.

\section{Literature Review}

Indoor localization, or positioning, techniques are well-studied topics in literature \cite{survey_positioning}. These methods contain the utilization of technologies such as camera systems, infrared, ultrasound, Wi-Fi, etc. The wireless technology-based localization systems among these technologies are divided into two parts as \emph{device-based} and \emph{device-free} \cite{survey_device}. Both of these approaches have their own advantages and drawbacks in terms of applicability, sensitivity, robustness, etc. \cite{survey_wireless}. Wi-Fi oriented localization techniques, in this matter, offer great opportunities, and a comprehensive study about this topic is given in \cite{survey_wifi}.

The first studies about CSI based localization appeared in \cite{fila_conf},\cite{fila_conf2},\cite{fila_journal}. These studies deploy a basic statistical approach to the localization problem. Later, various processing techniques were applied to CSI data such as wavelet transform \cite{wavelet}, visibility graphs \cite{visibilityGraph} and decision trees \cite{gbdt}.

The deployment of Deep Learning \cite{deeplearning}, or DNNs, to the CSI based localization schemes, emerged with the works \cite{deepfi_conf}, \cite{deepfi}. The basic fully connected neural network structures used in these works were advanced with \emph{convolutional neural networks} (CNNs) \cite{cnn}. While the general approach for these works is reshaping the data into spectrograms and feed it to a CNN architecture like done in \cite{CNN_pilc}, \cite{CNN_cifi} and, \cite{CNN_confi}, a 1-D convolution technique can be employed as done in \cite{1D_CNN}. Additionally, \emph{convolutinal deep autoencoder} structures are used for localization in \cite{autoencoder_rssi}. However, this approach employs only RSSI data for the localization task. There are also studies, like \cite{csi+rssi} and \cite{csi+rssi_outdoor}, where CSI and RSSI data are combined to improve performance. Furthermore, in \cite{vis}, a comprehensive analysis and visualization were done on the Deep Neural Networks employed for device-free localization tasks.

It is worth mentioning some of the studies of real-life mobile robot applications that use wireless localization techniques. In \cite{uav_rssi}, an RSSI-based approach has been followed to navigate a UAV. In \cite{uav_csi}, a UAV is automated to collect CSI data from the environment in an energy-optimal manner. Lastly, in \cite{agv_csi}, a CSI-based localization is applied to an automated guided vehicle.

\section{Scope of the Project}

This project mainly focuses on the utilization of CSI data for localization of a mobile unit. In this aspect, three main steps were followed throughout the project.

Firstly, a solid background about the basics of the wireless communication systems has been built, in order to develop certain expertise on the topic. This expertise is important to inspect the data planned to be acquired, both intuitively and quantitatively. In this aspect, the fundamental wireless channel models have been investigated and understood. In addition to that, a proper research has been done about the aforementioned OFDM technique. That is how, possible complications during the practical part of the project due to the lack of knowledge have been avoided.

As the second step of the project, the infrastructure and a proper routine for the data collection process have been built. The building of the infrastructure required both software and hardware designs on the wireless communication units. In this aspect, two wireless router units with proper chipsets are utilized as transmitter and receiver. The required software modifications and integrations have been done for these routers. Also, in order to collect more reliable CSI data, certain precautions were taken on the routers hardware-wise. After validating that the acquired CSI data is healthy, the environment where the localization process is held has been prepared. Lastly, the data collection process has been initiated with a routine that avoids the data imbalance between the coordinates of interest.

\chapter{BASICS OF WIRELESS COMMUNICATION}

Communication is a process by which information is exchanged between individuals through a common system of symbols, signs, or behavior through a certain channel \cite{merweb}. A receiver, a transmitter and a channel must be present for the process to take place. The transmitter transmits the signals that it loads information to the receiver via the channel. The information to be sent is exposed to certain random effects until it reaches the transmitter. These effects cause deterioration in the signals carrying the information. These effects affect the information signal multiplicative or additively. Noise is the name given to irregular and random disturbances that accompany but do not belong to it during the transmission of a signal.

\begin{figure}[!h]
    \centering
    \vspace{5mm}
    \includegraphics[scale = 0.4]{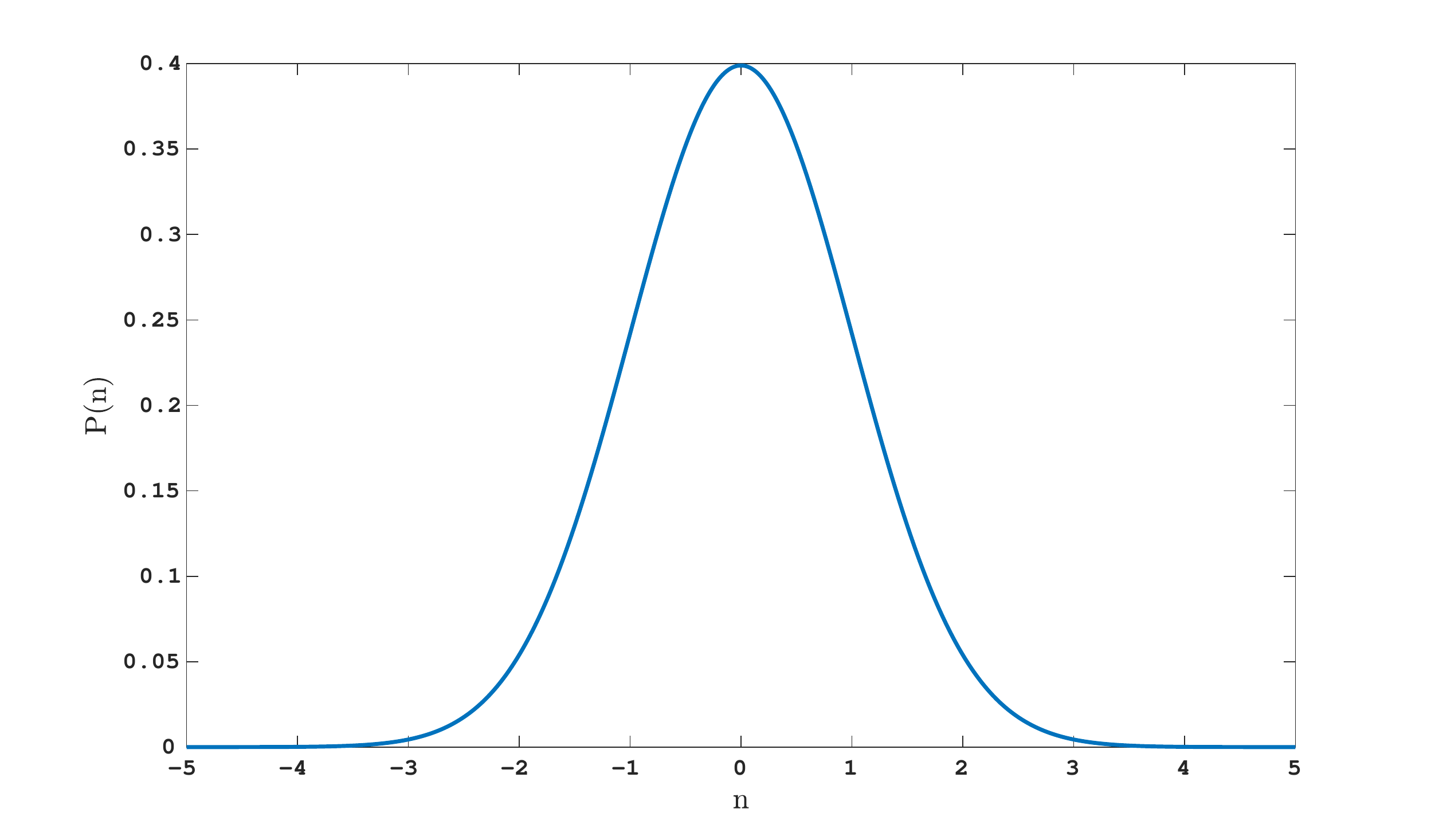}
    \vspace{5mm}
    \caption{Zero mean and unit variance Gauss PDF.}
    \label{fig:gausspdf}
\end{figure}

The time domain representation of noise, \emph{n(t)}, can be given as a Gaussian probability density function (PDF) with zero mean and $\sigma^2$ variance, i.e.
\begin{equation}
    n(t) \sim \mathcal{N}(n(t);0,\sigma^2) = \frac{1}{\sqrt{2\pi\sigma^2}}e^{-\frac{n(t)^2}{2\sigma^2}}.
\end{equation}
PDF of a Gauissan with zero mean and unit variance is given in Fig. \ref{fig:gausspdf}. This type of noise is called white Gaussian noise, because it contains all the frequency components equally and the amplitude of it distributes as Gauss in time domain.

There are many random channel models that demonstrate the behavior of the channel in communication systems. Some of these can be listed as
\begin{itemize}
    \item Additive White Gaussian Noise (AWGN) channel
    \item Rayleigh fading channel
    \item Rician fading channel
    \item Nakagami-$m$ fading channel.
\end{itemize}
These listed channel models are investigated in the following sections.

\section{AWGN Channel}

\begin{figure}[!h]
    \centering
    \vspace{0mm}
    \includegraphics[scale = 0.45]{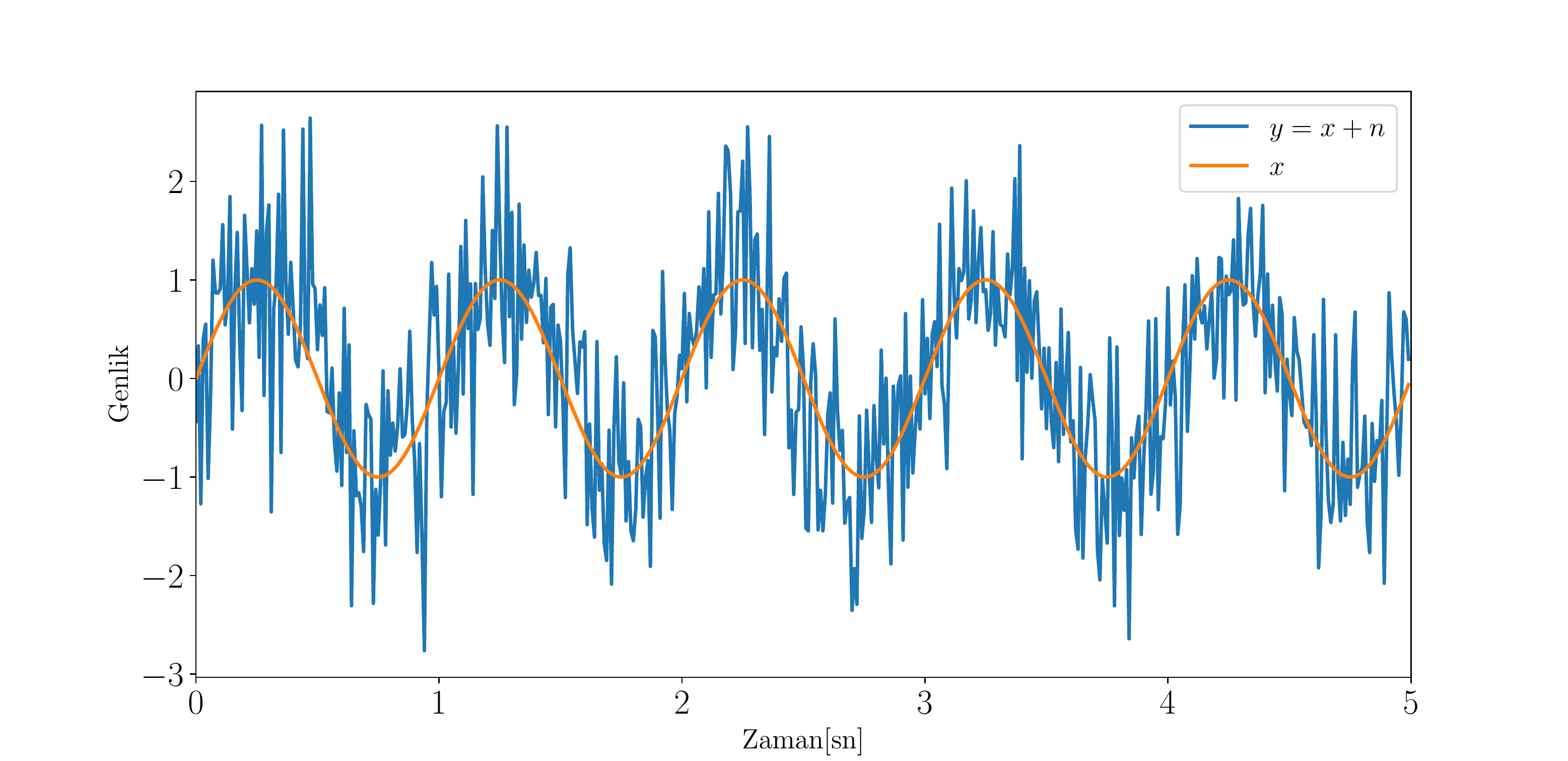}
    \vspace{5mm}
    \caption{The reception of the transmitted signal $x$ in the receiver as $y$ under AWGN channel.}
    \label{fig:gauss}
\end{figure}

In a communication system under AWGN channel effects, the received signal can be written as
\begin{equation}
    y = x + n
    \label{eqn:awgn}
\end{equation}
where $x$ is the transmitted signal which carries information and $n$ is AWGN. Here, the time indices are omitted since this relation applies for all time instances. It can be seen a sinusoidal signal corrupted by an AWGN channel in Fig. \ref{fig:gauss}.

\newpage

\section{Fading Channel Concept and Fading Channel Parameters}

In wireless communication, electromagnetic waves emitted from the transmitting antenna scatter in different directions and interact with the environment. As a result, the antenna in the receiver receives the information signals carried by the electromagnetic waves and scattered in the environment, together with different fading and phase shifts.

\begin{figure}[!h]
    \centering
    \vspace{5mm}
    \includegraphics[scale = 0.7]{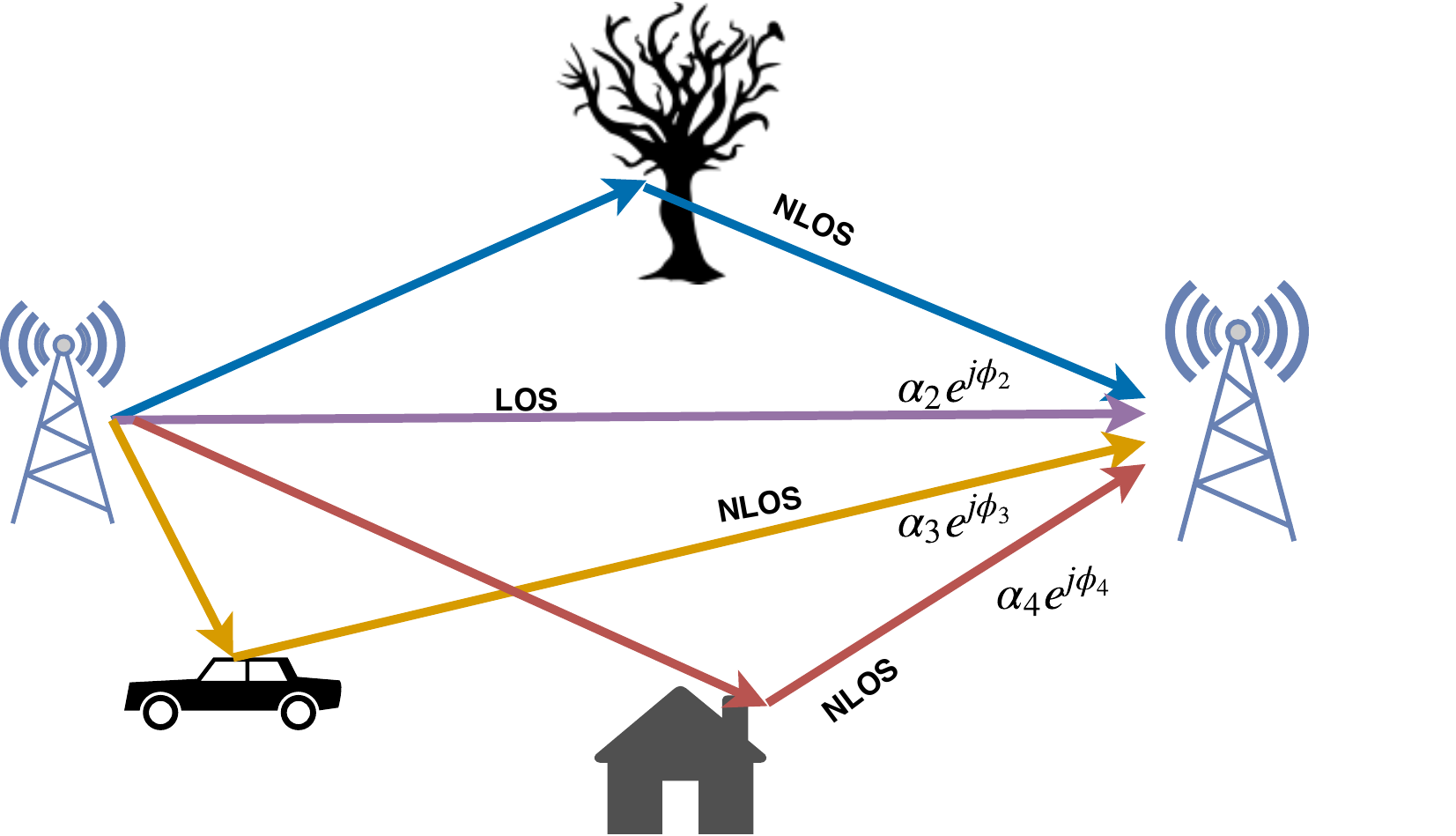}
    \vspace{5mm}
    \caption{Scattering of the electromagnetic wave sent from the transmitter and reaching the receiver in different amplitude-phases.}
    \label{fig:fadingenv}
\end{figure}

In Fig. \ref{fig:fadingenv}, a fading channel example is given visually. As can be seen in the figure, the signal sent from the transmitter comes directly to the receiver (line-of-sight, LOS), as well as reaching the receiver by spreading from the environment (non-line-of-sight, NLOS).

Let the information signal sent from the transmitter is shown as $x$ under the fading channel. The received signal in noiseless case can be given as
\begin{equation}
    y = x\alpha_1e^{j\phi_1} + x\alpha_2e^{j\phi_2} + ... + x\alpha_Ne^{j\phi_N} = x\lbrace \alpha_1e^{j\phi_1} + \alpha_2e^{j\phi_2} + ... +\alpha_Ne^{j\phi_N}\rbrace
\end{equation}
when $N$ copies of $x$ reaches the receiver with different amplitudes and phases, assuming that they have the same amount of delay. Here, $\alpha_i$ and $\phi_i$ represent the loss and phase shift of the $i$th path. This expression can be shortened as
\begin{equation}
    y = x\sum_{i=1}^{N}\alpha_ie^{j\phi_i}
\end{equation}
In the noiseless case, the relationship between the $y$ signal and the $x$ signal to the receiver is an environment-related multiplier. Therefore, the fading channels generally affect the information sign in a multiplicative way. A more generalized verison of this expression can be given as
\begin{equation}
    y = hx + n.
\end{equation}
Here $h$ represents the channel coefficient that determines the effect of the channel and written as
\begin{equation}
    h = \sum_{i=1}^{N}\alpha_ie^{j\phi_i}.
\end{equation}
Therefor, $h$ must be a random variable with statistical properties to model randomness of the channel.

Some of the notions that forms the characteristics of the fading channels can be listed as
\begin{itemize}
    \item Attenuation
    \item Fading
    \item Propagation delays
    \item Doppler frequency shifts.
\end{itemize}
Summarized information about these notions will be given in the following subsections.

\subsection{Attenuation}

The information signal intended to be transmitted in the communication systems is released to the channel with a certain power level. During the transmission of the information signal on the channel, there is a loss in the power of this signal. As a matter of fact, the power perceived by the receiver is at a lower level than the power of the information signal transmitted from the transmitter to the channel. This event is called attenuation.

\newpage

\subsection{Fading}

Especially in wireless communication, the notion of fading dramatically affects the communication performance. In radio communication, when the signal emitted from the transmitter is released to the environment, it is reflected from many objects and reaches to the receiver by scattering. Information signals reflected and scattered in different ways are received additively in the receiver. Due to the randomness in the amplitude and phases of the information signals coming to the receiver, this sum may affect the information signal obtained in the receiver as constructive or destructive. The situation that occurs as a result of the destructive effect of this randomness on the information signal received from the transmitter is called fading.

\subsection{Propagation Delays}

\begin{figure}[!h]
    \centering
    \vspace{5mm}
    \includegraphics[scale = 0.42]{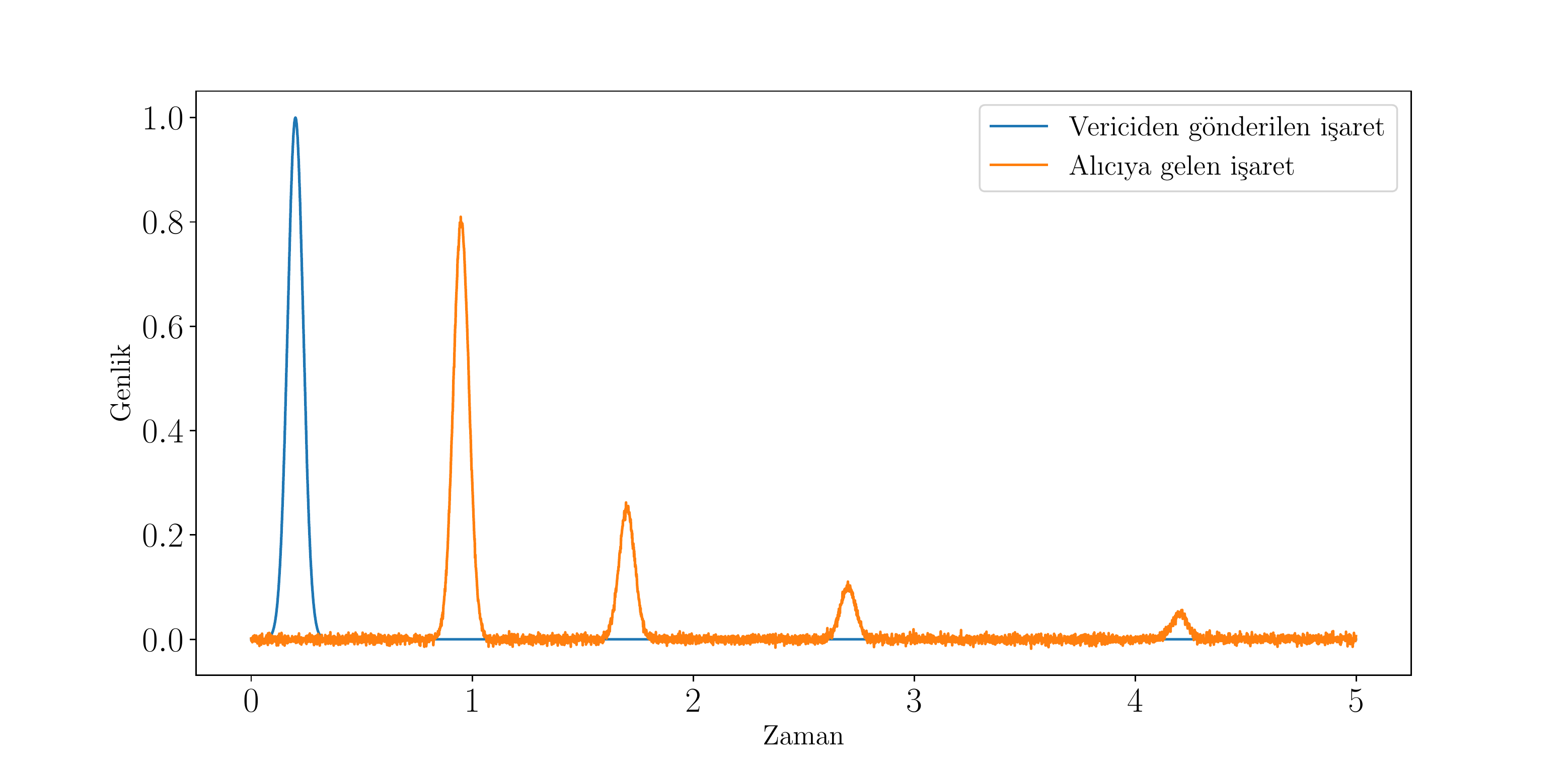}
    \vspace{5mm}
    \caption{Gaussian pulse sent from the transmitter reaches the receiver at different times and amplitudes.}
    \label{fig:delayspread}
\end{figure}

In wireless communication, the information signal that transits from the transmitter to the channel reaches to the receiver by traveling at different distances in the environment. For this reason, the information signal reaches the receiver in the form of different delays and the sum of the attenuations. This event is referred to as the propagation delay. In Fig. \ref{fig:delayspread}, it is seen that a Gaussian pulse sent from the transmitter reaches the receiver with different delays, different attenuations and additive noise.

\subsection{Doppler Frequency Shift}

\begin{figure}[!h]
    \centering
    \vspace{5mm}
    \includegraphics[scale = 0.55]{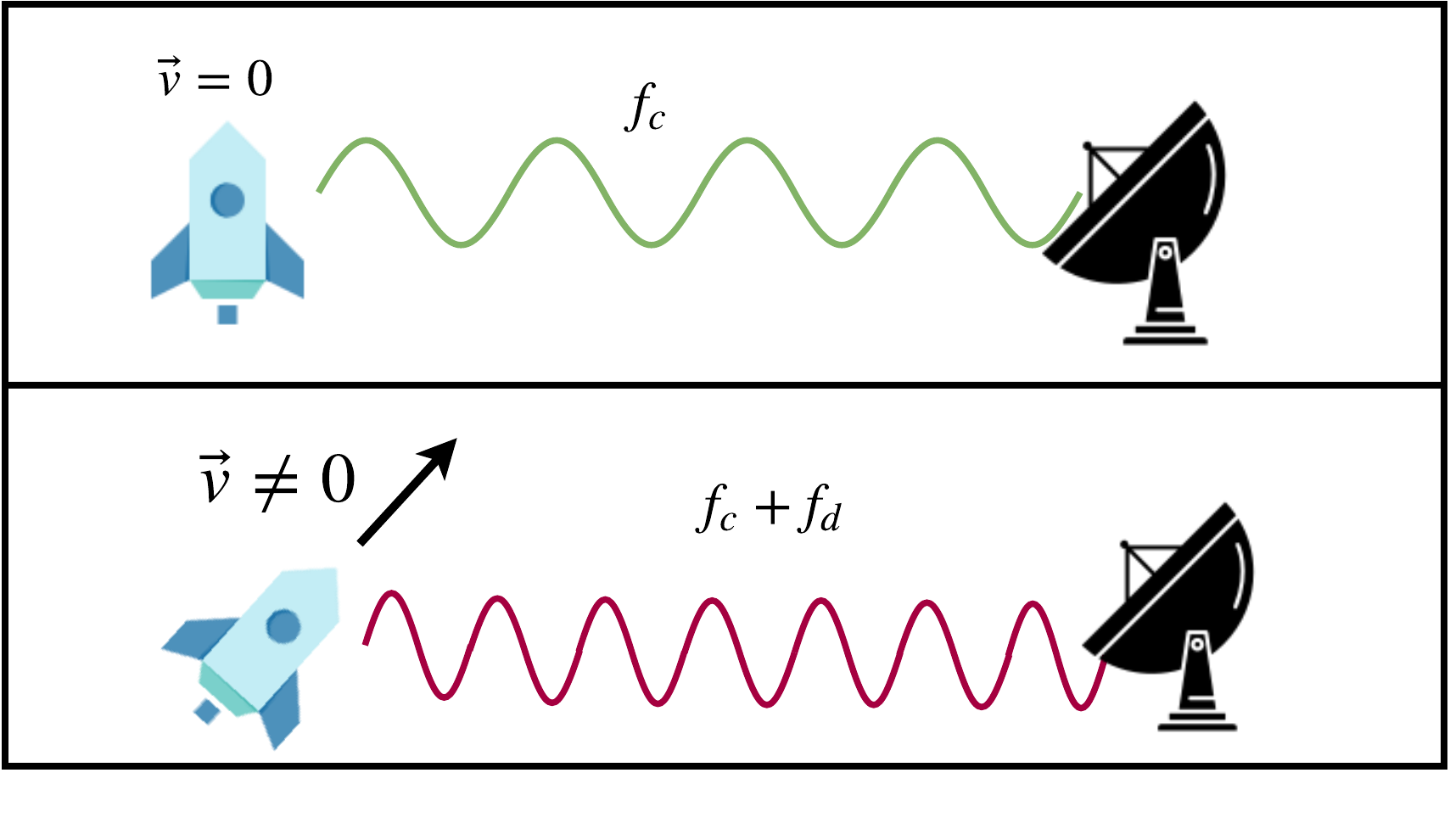}
    \vspace{5mm}
    \caption{The detection of electromagnetic wave with a frequency of $f_c$ as $ f_c + f_d $ while the transmitter has the relative speed.}
    \label{fig:dopplershift}
\end{figure}

Suppose a pure sinusoidal with a frequency of $f_c$ is emitted by a transmitter. When the radiating signal reaches the receiver that has the speed relative to the source, a change in the frequency of this sinusoidal signal is observed around $f_c$. The relationship between shifts on the frequency of $ f_c $ and the relative speed is expressed as
\begin{equation}
    f_d = f_c\frac{vcos\theta}{c}
\end{equation}
where $f_d$, $v$, $c$ and $\theta$ represent the frequency shift, the relative speed between transmitter and receiver, the propagation speed of the signal in the environment and the angle between the relative velocity vector and the propagation vector, respectively. This event is called the Doppler frequency shift and is depicted in Fig. \ref{fig:dopplershift}. Although the receiver or transmitter does not have different speeds compared to each other, changes in the environment are also sufficient enough for observing the Doppler frequency shift event.

\section{Fading Channel Models}

In the previous section, the channel effects used in the characterization of fading channels are given. In this section, the statistical properties of certain fading channel models are given. In the analysis, the results were obtained by assuming that the channel is a stationary random process.

\newpage

\subsection{Rayleigh Channel Model}

\begin{figure}[!h]
    \centering
    \vspace{5mm}
    \includegraphics[scale = 0.4]{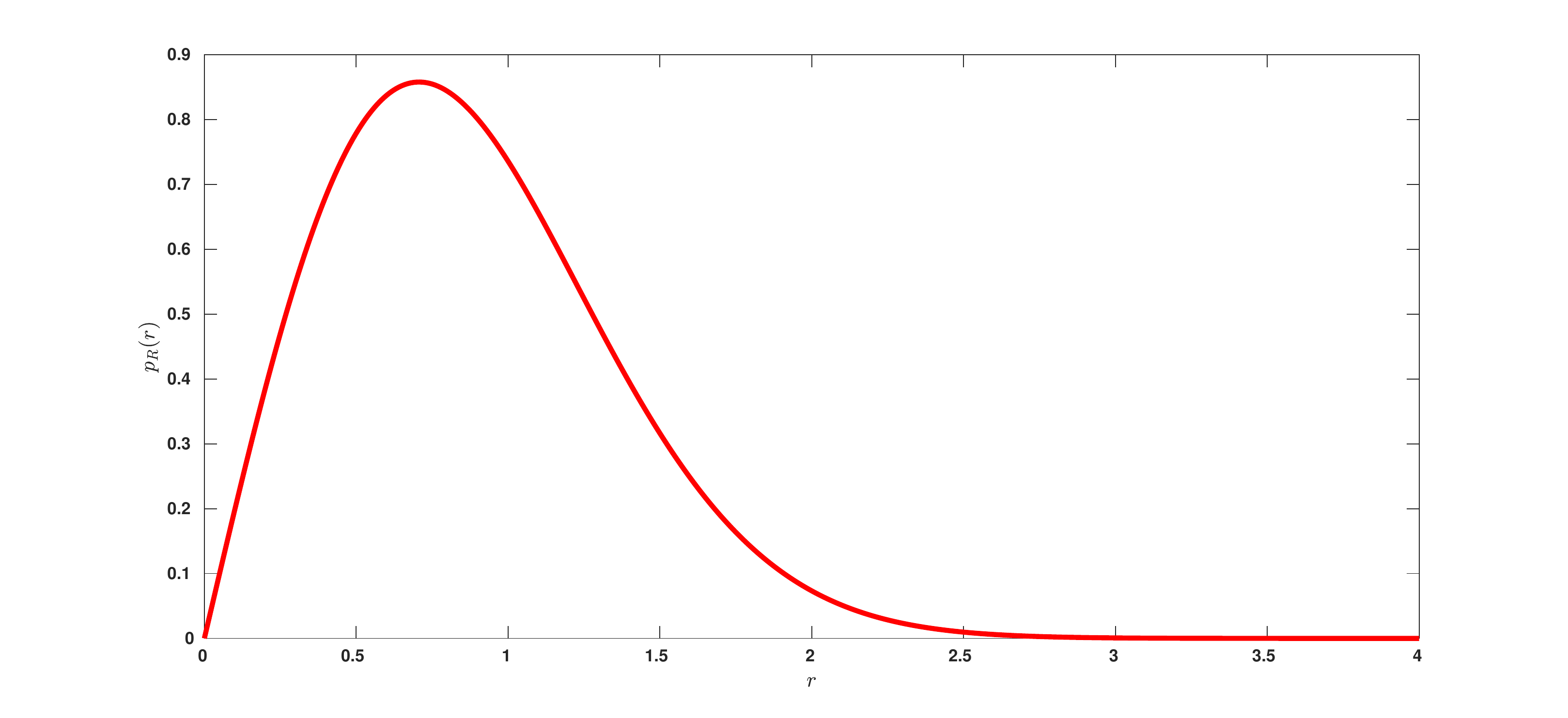}
    \vspace{5mm}
    \caption{PDF of Rayleigh distribution.}
    \label{fig:rayleighdist}
\end{figure}

The Rayleigh channel model is the channel model in which there is no direct path between the receiver and the transmitter, the signal coming to the receiver is the result of reflections and there are infinite reflections. The impulse response of the channel in time domain is given as the summation of
\begin{equation}
    h(t) = \sum_{i=1}^{\infty}\alpha_i\delta(t-\tau_i).
\end{equation}
This expression corresponds to
\begin{equation}
    h = \sum_{i=1}^{\infty}\alpha_ie^{j\phi_i} = \sum_{i=1}^{\infty}\alpha_icos(\phi_i) -j\sum_{i=1}^{\infty}\alpha_isin(\phi_i)
\end{equation}
in frequency domain. The real($a$) and imaginary($b$) parts of $h$ are
\begin{equation}
    a = \sum_{i=1}^{\infty}\alpha_icos(\phi_i) \quad b = \sum_{i=1}^{\infty}\alpha_isin(\phi_i),
\end{equation}
so that $h$ can be written as
\begin{equation}
    h = a +jb.
\end{equation}
If $a$ and $b$ are normalized to have Gaussian PDFs with means of zero and variances of $\frac{1}{2}$, the PDF expressions for $a$ and $b$ can be written as
\begin{equation}
    p_A(a) = \frac{1}{\sqrt{\pi}}e^{-a^2}, \quad p_B(b) = \frac{1}{\sqrt{\pi}}e^{-b^2}.
\end{equation}
Since $a$ and $b$ random variables are independent and distributed as Gaussians, they are uncorrelated. Therefore, the joint PDF of these random variables can be given as
\begin{equation}
    p_{AB}(a,b) =p_A(a)p_B(b) = \frac{1}{\pi}e^{-\left(a^2 + b^2\right)}.
\end{equation}
Using this information, the statistical properties of $h = a +jb$ can be calculated.
If the amplitude and the phase of $h$ is represented as $r$ and $\theta$, respectively, $h$ can be expressed as
\begin{equation}
    h  \triangleq re^{j\theta}.
\end{equation}
Therefore, the relation between $a$, $b$, $r$ and $\theta$ is given as
\begin{equation}
    r = \sqrt{a^2 + b^2} \quad \quad \theta = tan^{-1}\left(b/a\right).
\end{equation}
The joint PDF of $r$ and $\theta$ can be calculated in terms of $p_{AB}$ as
\begin{equation}
    p_{R\Theta}(r,\theta) = p_{AB}(a,b)\mid J_{AB} \mid .
\end{equation}
Here, $\mid J_{AB} \mid$ represents the determinant of the Jacobian matrix and calculated as
\begin{equation}
    \mid J_{AB} \mid = det
    \begin{bmatrix}
        \frac{\partial a}{\partial r}      & \frac{\partial b}{\partial r}      \\
        \frac{\partial a}{\partial \theta} & \frac{\partial b}{\partial \theta}
    \end{bmatrix} = r .
\end{equation}
After the required algebraic computations, one can obtain the joint distribution as
\begin{equation}
    p_{R\Theta}(r,\theta) = \frac{r}{\pi}e^{-r^2}.
\end{equation}
In order to achieve the expression for $p_{R}(r)$, the integration
\begin{equation}
    p_R(r) = \int_{-\pi}^{\pi}  \frac{r}{\pi}e^{-r^2} d\theta
\end{equation}
is done and the result is obtained as
\begin{equation}
    p_R(r) = 2re^{-r^2}, \quad r\geq 0
\end{equation}.
$p_\Theta(\theta)$ expression is obtained in the same manner as
\begin{equation}
    p_\Theta(\theta) = \frac{1}{2\pi}, \quad -\pi<\theta<\pi.
\end{equation}
As it can be seen, the phase shift occurs with the same probability for every phase. Since the statistical properties of the channel coefficient amplitude are in the form of Rayleigh distribution (Fig. \ref{fig:rayleighdist}), this channel model is named as Rayleigh fading channel.

\subsection{Rician Channel Model}

In addition to the Rayleigh channel model, there is an additional way between the transceiver and the Rician channel model. Let the amplitude of the direct transmission between the receiver and the transmitter be $\sqrt{D^2}$ so that the channel coefficient can be expressed as
\begin{equation}
    h = D + a +jb.
\end{equation}
$a$ and $b$ are random Gaussian variables with zero mean and $\frac{1}{2}$ variance that determine the randomness of the channel. It is clear that $E\left[\mid h \mid ^2 \right]$ will be greater than 1 for $ h = D + a + jb $. So if $h$ is normalized to $E\left[\mid h \mid ^2 \right] = 1$, it is written as
\begin{equation}
    h \triangleq \frac{D + a +jb}{\sqrt{D^2 + a^2 + b^2}}.
\end{equation}
PDF of the amplitude of the channel coefficent $h$ under Rician channel is given as
\begin{equation}
    p_R(r) = \frac{r}{\sigma_s^2}e^{-\frac{n^2+D^2}{2\sigma_s^2}}I_0\left(\frac{rD}{\sigma_s^2}\right), r\geq0.
    \label{eqn:rician0}
\end{equation}
Here, $r$ and $\sigma_s^2$ are used to represent $\mid h\mid$ and variance, respectively. $I_0(.)$, on the other hand, is the zeroth order first kind modified Bessel function.

Let us define the ratio between the powers of the deterministic and probabilistic parts of the channel between the transmitter and the receiver as
\begin{equation}
    K = \frac{D^2}{2\sigma_s^2},
\end{equation}
so that the expression (\ref{eqn:rician0}) is rewritten as
\begin{equation}
    p_R(r) = 2r(1 + K)e^{-K-(1+K)r^2}I_0\left(2r\sqrt{K(K+1)}\right), \quad r\geq0
    \label{eqn:rician}.
\end{equation}

\subsection{Nakagami-$m$ Channel Model}

\begin{figure}[!h]
    \centering
    \includegraphics[scale = 0.450]{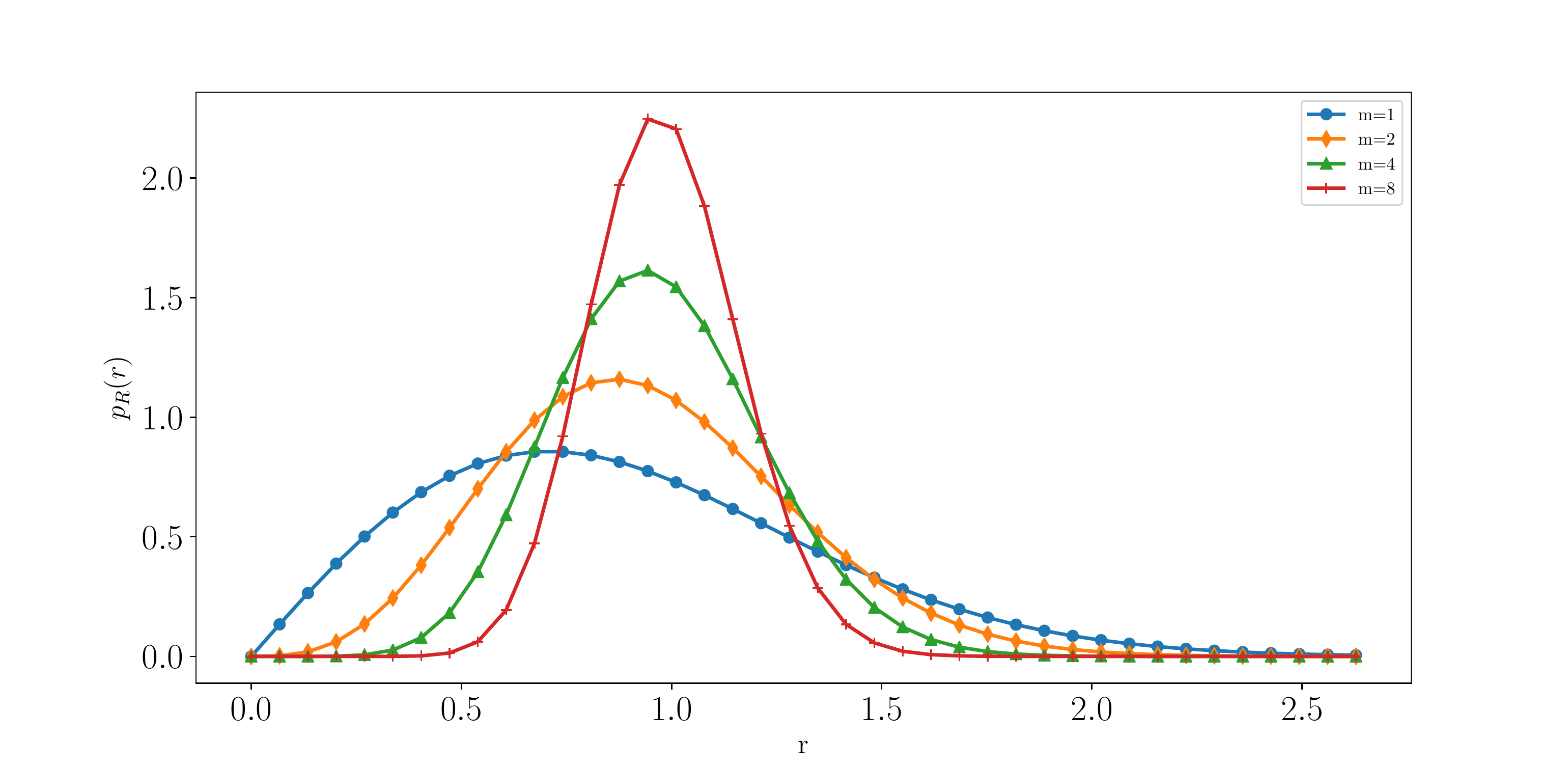}
    \vspace{5mm}
    \caption{PDF of Nakagami-$m$ channel for different values of $m$.}
    \label{fig:nakagamidist}
\end{figure}

Rayleigh and Rician channel models are used to model NLOS and LOS channels, but they are not sufficient to model all channel types encountered in practical applications. For example, although the Rayleigh channel model is an NLOS channel model, in practice, not all NLOS channels show the same behavior. So the distribution that models all these NLOS channels needs some freedom to determine the character of the channel. Nakagami-$m$ is a channel model with two degrees of freedom and derived from Gamma PDF. One of these freedoms is $m$ and the other one is $\Omega$. The PDF of Nakagami-$m$ channel is expressed as
\begin{equation}
    p_R(r) = \frac{2m^m}{\Omega^m\Gamma(m)}r^{2m-1}e^{-\frac{n^2m}{\Omega}}.
\end{equation}
The Nakagami-$m$ channel model can be transformed into Rayleigh, Rician and AWGN channels by manipulating its two parameters. Thus, a Rician distribution free from the modified Bessel function, which cannot be taken out analytically in (\ref{eqn:rician}), can be obtained. In Fig. \ref{fig:nakagamidist}, while $\Omega =1$, Nakagami-$m$ distributions are given for different $m$ values. As it can be seen, as $m$ increases, PDF gets dense around 1 and it turns into AWGN channel.

\section{OFDM Technique}

In the last sections, the statistical characteristics of various channel models and underlying reasons them to occur has been given. While explaining these concepts, it has been assumed that the communication is conducted by signals of frequency $f_c$. Also, even though it was mention in the underlying physical notions of the channel, the propagation delay effect has been omitted for the rest of the channel derivations. In other words, it was assumed that all of the scattered signals reach the receiver at the same time. It means that the channel impulse response for this case is a single impulse whose magnitude is a random variable depends on the channel model. Therefore, the frequency spectrum of the channel is flat for all frequencies.

\begin{figure}[!h]
    \centering
    \vspace{3mm}
    \includegraphics[scale = 0.40]{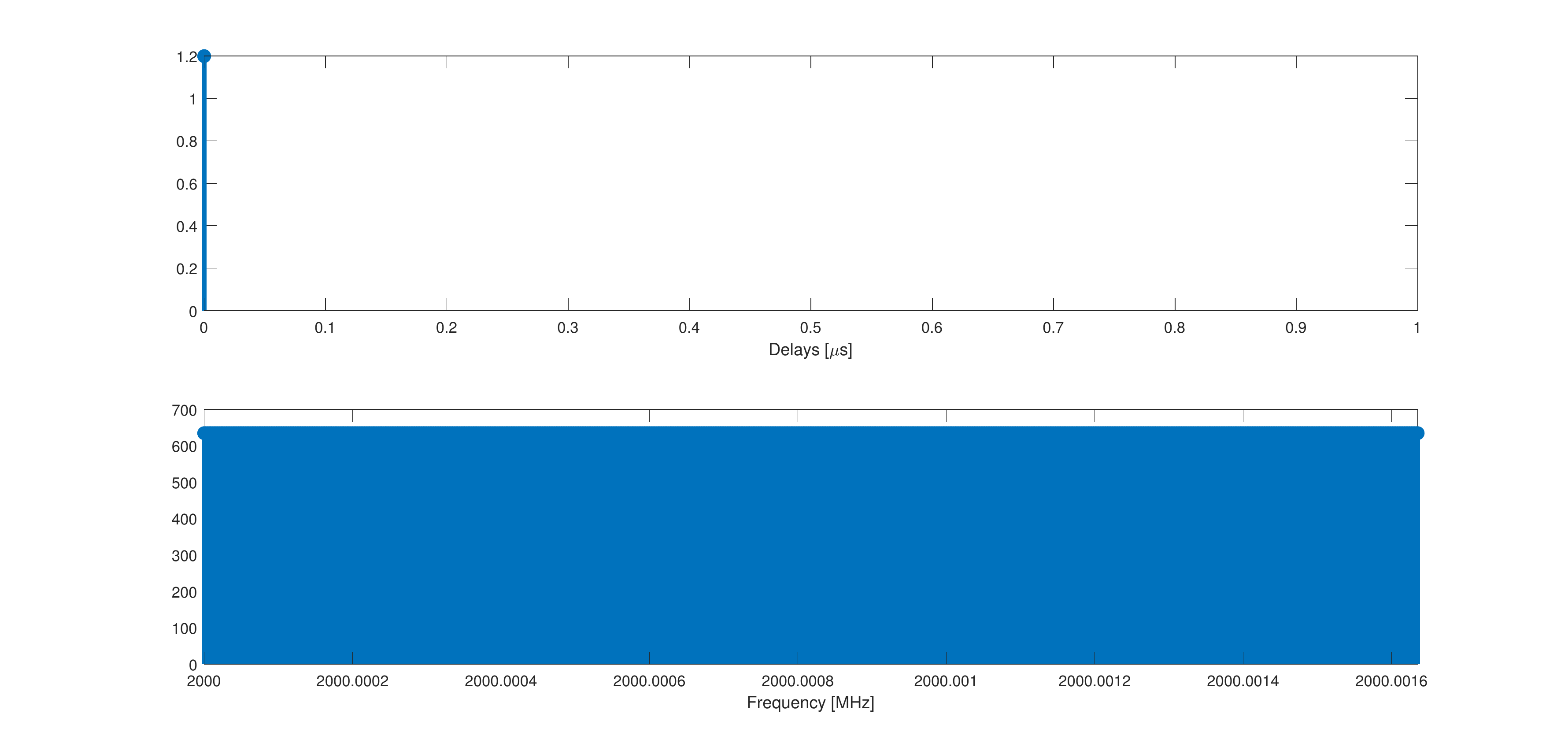}
    \vspace{3mm}
    \caption{An instance of a unit power Rayleigh distributed single tap channel delay profile and frequency spectrum.}
    \label{fig:singletap}
\end{figure}

\begin{figure}[!h]
    \centering
    \vspace{3mm}
    \includegraphics[scale = 0.40]{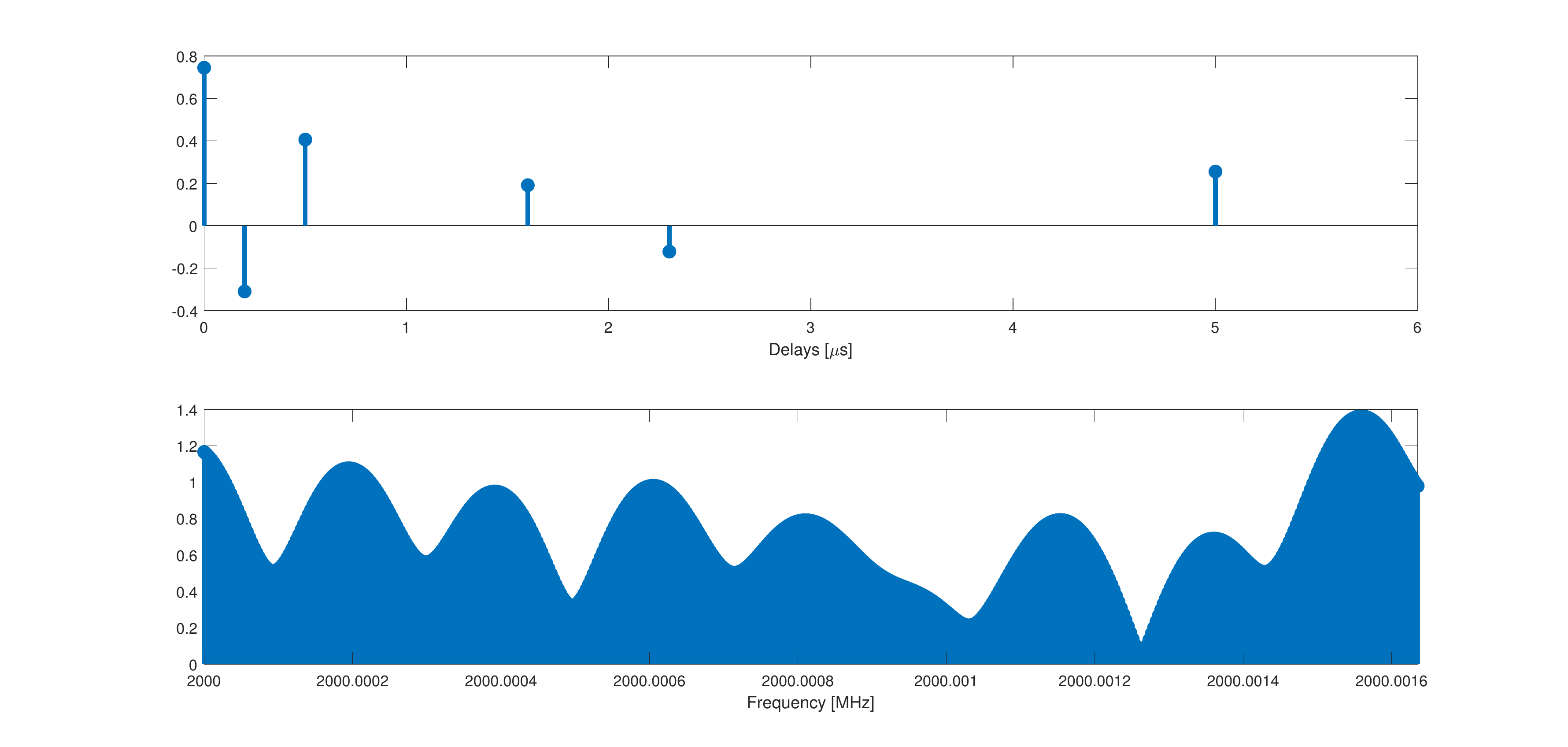}
    \vspace{3mm}
    \caption{An instance of a Rayleigh distributed mult-tap channel delay profile and frequency spectrum.}
    \label{fig:multitap}
\end{figure}

The assumption of \emph{single tap} channels are far from practice. In realistic scenarios, the channel is assumed to be \emph{multi-tap}, i.e. receiver receives multiple copies of the signal with different magnitudes and phases. Thus, the assumption of flatness of the channel spectrum is no longer valid. This concept can be shown with a basic Fourier transform. In Fig. \ref{fig:singletap} an instance of a single tap channel is given. It can be seen that the Fourier transform of the delay profile gives a flat frequency spectrum. In Fig. \ref{fig:multitap}, on the other hand, a multi-tap delay profile with different powers for each tap is given and it results in a highly distorted frequency spectrum. This kind of channels are named as \emph{frequency selective} channels.

In practice, in order to get rid of the distortion in the frequency spectrum, the receiver should estimate the channel frequency response and compensate the unwanted distortion. However, this process is very ineffective and sometimes impossible to implement for single-carrier communication systems. Therefore, it is desired to use narrow-band signals for single-carrier systems, so that the frequency spectrum can be approximated as flat in the narrow frequency band. However, narrower information signal spectrum means slower data rates.

\begin{figure}[!h]
    \centering
    \vspace{3mm}
    \includegraphics[scale = 0.40]{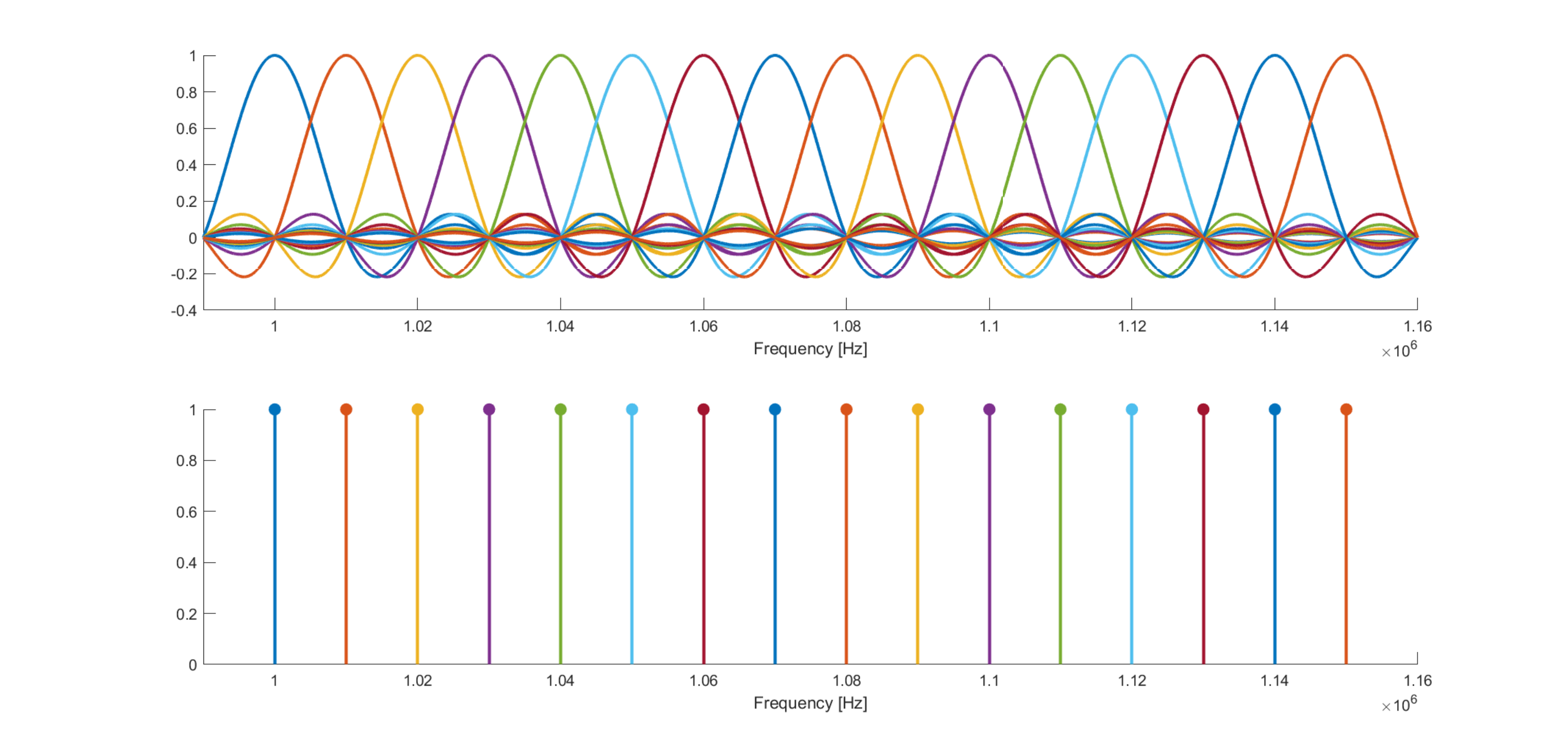}
    \vspace{3mm}
    \caption{Illustration of an OFDM carrier spectrum.}
    \label{fig:ofdm}
\end{figure}

One way to cope with the trade-off between the flatness of the spectrum and the data rate is the technique called OFDM. The basic principle of OFDM is designing the information signal in the frequency domain and transmitting the signal after taking the inverse fast Fourier transform (IFFT) of it. The design in the frequency domain is basically stacking lots of narrow-band information signals in an orthogonal way, i.e. they do not interfere with each other. This orthogonality is satisfied by the \emph{sinc} shaped \emph{subcarriers}. The general view of an OFDM signal carrier spectrum is given in Fig. \ref{fig:ofdm}. Here, the signal contains 16 subcarriers. One can notice that on the peak points of each subcarrier, the rest of the subcarriers take the value of zero. It also can be seen from the same figure that taking samples from exactly the middle of the subcarriers results in nothing but the corresponding subcarrier amplitude. Thus, this technique avoids the event \emph{inter-channel interference}.

The aspect of flattening the channel of OFDM brings the advantage of inspect the frequency selective channel itself. If the transmitted symbols are known by the receiver apriori, i.e. transmitter sends pilot symbols, the receiver can obtain the channel coefficients corresponding to every subcarrier. This collection of channel coefficients is named as \emph{Channel State Information} or CSI. In practice, CSI is used to compensate the distortion which comes with the delay profile as it was mentioned before. However, in this project, CSI is used to characterize the channel so that the frequency response of the channel can be matched with the current location of the transmitter.
\chapter{EXPERIMENT SETUP}

In this chapter, the experiment setup and the CSI data collection process is given. All of the experiments were conducted in Istanbul Technical University Artificial Intelligence and Intelligent Systems Laboratory.

\section{Hardware and Software Infrastructre}

CSI collection is not available for every commercial wireless device, unfortunately. This feature became available for the user with IEEE 802.11n standardization. However, not every chipset producer included this feature in their product. Two main chipset producers manufacture sufficient chipsets in this aspect are Intel and Qualcomm. Even though Intel chipsets are more common in literature for CSI based studies, it is not feasible to acquire the proper chipsets in Turkey. Therefore, Qualcomm branded chipsets are preferred in this project. However, the same infeasibility remains for the standalone Qualcomm chipsets. Luckily, there are wireless routers in the market equipped with required chipsets \footnote{\url{https://github.com/xieyaxiongfly/OpenWRT_firmware}}.

\begin{figure}[!h]
    \centering
    \vspace{5mm}
    \includegraphics[scale = 0.5]{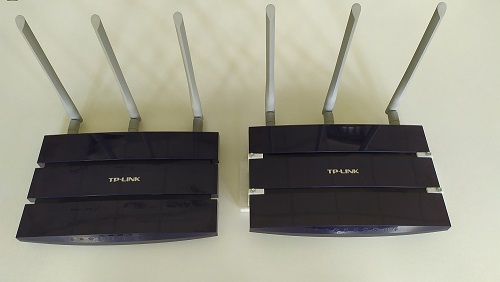}
    \vspace{5mm}
    \caption{Transmitter and receiver TL-WR1043ND routers.}
    \label{fig:routers}
\end{figure}

It is decided to use the TL-WR1043ND model of TP-Link wireless routers both for the transmitter and the receiver. These routers are equipped with Qualcomm Atheros QCA9563 chipsets which are appropriate for CSI data acquisition. In Fig. \ref{fig:routers}, it can be seen both of the routers used in the project. Further hardware information about these routers are provided in the Internet \footnote{\url{https://openwrt.org/toh/tp-link/tl-wr1043nd}}. This information is crucial because the integration of bulky routers to mobile robots may not be possible. Therefore, this integration should be done by the designer and it requires deep hardware knowledge to implement.

\begin{figure}[!h]
    \centering
    \vspace{5mm}
    \includegraphics[scale = 0.6]{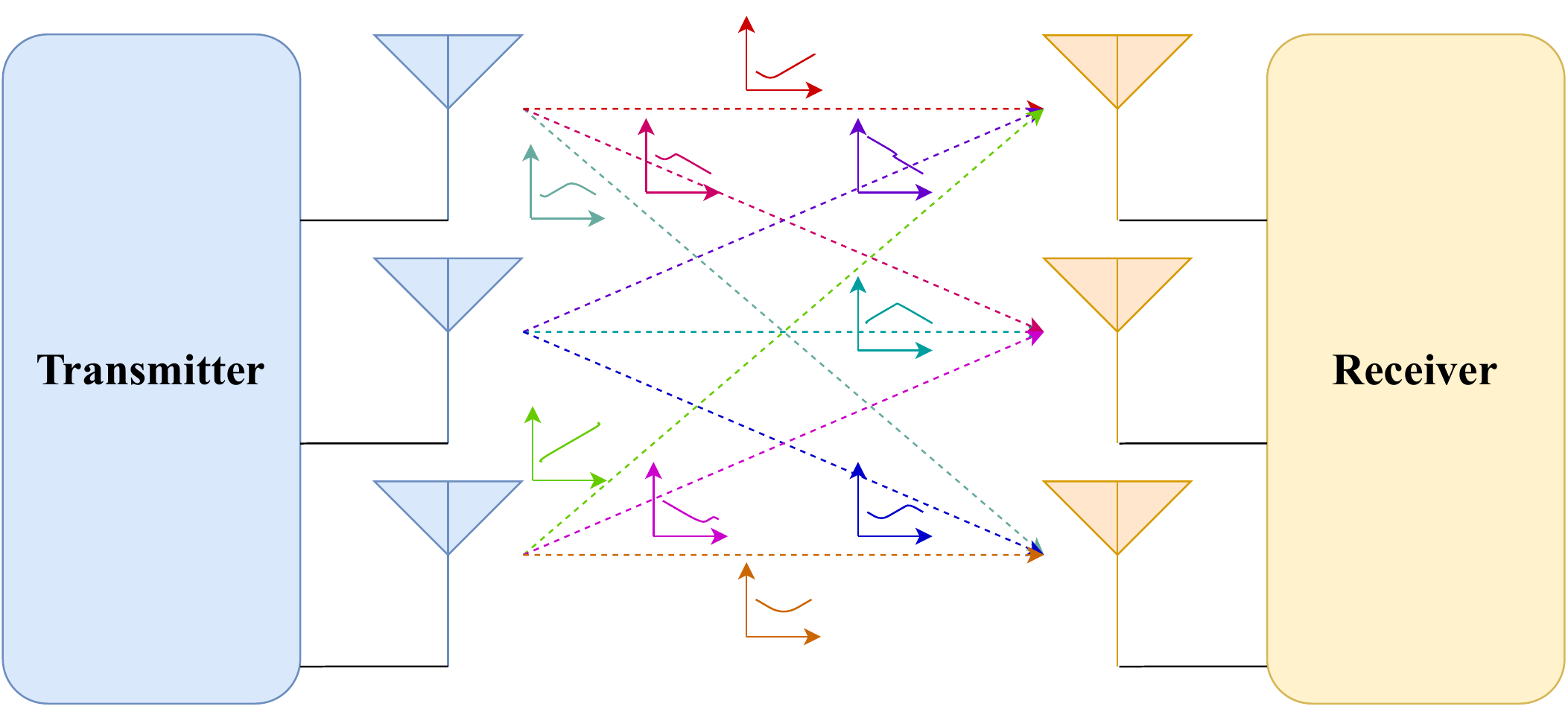}
    \vspace{5mm}
    \caption{9 communication links for 3x3 antenna combination and related (figurative) channel spectrums.}
    \label{fig:3x3}
\end{figure}

The main reason to choose these routers for this project is that they are equipped with 3 antennas. This is important because each antenna pair crate a link between them and each link is exposed to different channel effects, in theory. Therefore, in every time instance, there occur 9 different channels between the transmitter and the receiver. This event is depicted in Fig. \ref{fig:3x3} (best viewed in color). These 9 different channels are later used as the \emph{fingerprints} of the corresponding locations.

\begin{figure}[!h]
    \centering
    \vspace{5mm}
    \includegraphics[scale = 0.5]{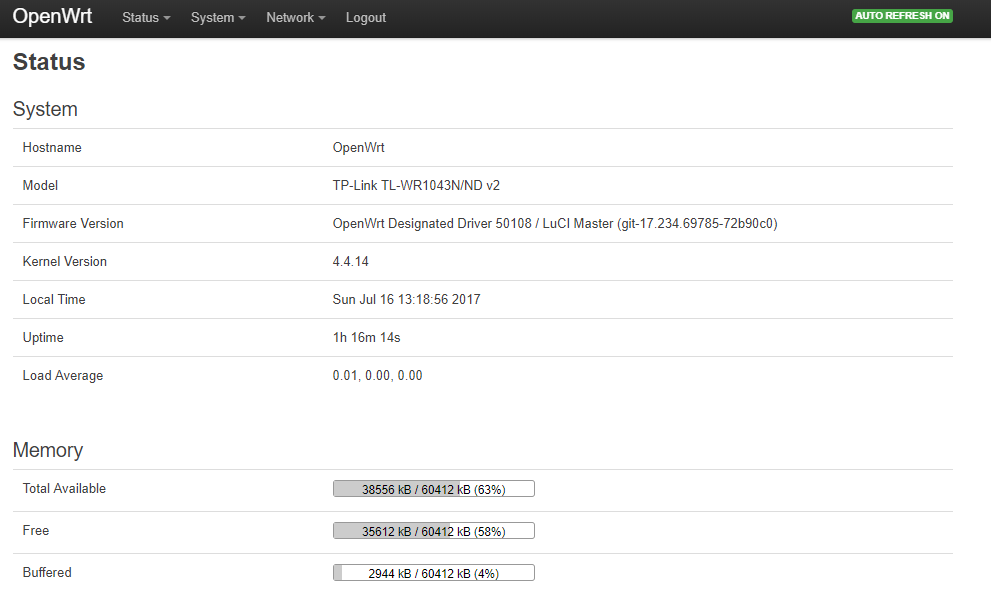}
    \vspace{5mm}
    \caption{Homepage of LuCI user interface.}
    \label{fig:luci_home}
\end{figure}

\begin{figure}[!h]
    \centering
    \vspace{5mm}
    \includegraphics[scale = 0.8]{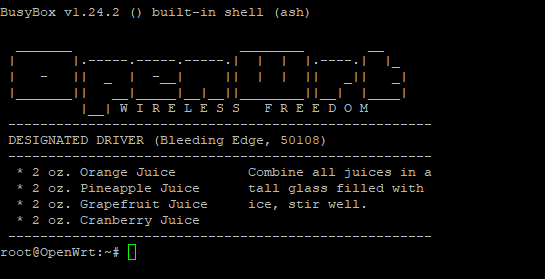}
    \vspace{5mm}
    \caption{Terminal of OpenWrt.}
    \label{fig:openwrt_ssh}
\end{figure}

Even though the wireless router chipsets allow us to collect and save CSI data, the interaction with the routers requires additional software arrangments. The first step of these arrangments is converting the commercial interface with a more developer-friendly interface called LuCI. This user interface is used to configure OPEN Wireless Router (OpenWrt) which is an open-source Linux based embedded operating system to route network traffic. OpenWrt offers lots of various software packages for networking and highly used both in industry and academia. The screen views of the homepage of LuCI and the terminal of OpenWrt are given in Fig. \ref{fig:luci_home} and Fig. \ref{fig:openwrt_ssh}.

LuCI interface is used for higher-level assignments to routers. Firstly, Internet Protocol (IP) addresses are assigned to each router. Then, these routers are included in a wireless network so that they can communicate with each other. Additionally, the wireless communication band and transmit powers are adjusted. That is how, these aspects are kept constant during the experiments.

\begin{figure}[!h]
    \centering
    \vspace{5mm}
    \includegraphics[scale = 0.9]{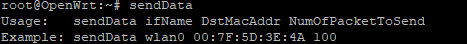}
    \vspace{5mm}
    \caption{Basic usage of command \texttt{sendData} from the OpenWrt terminal.}
    \label{fig:sendData}
\end{figure}

\begin{figure}[!h]
    \centering
    \vspace{5mm}
    \includegraphics[scale = 0.9]{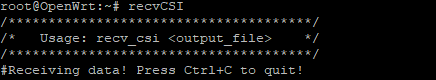}
    \vspace{5mm}
    \caption{Basic usage of command \texttt{recvCSI} from the OpenWrt terminal.}
    \label{fig:recvCSI}
\end{figure}

The acquisition of CSI requires the installation of an adequate CSI collecting software appropriate to OpenWrt. Since the routers contain Qualcomm Atheros chipsets, Atheros CSI Tool \cite{csitool} is appropriate to use. This tool can be downloaded from the Internet for a wide range of router models \footnote{\url{https://github.com/xieyaxiongfly/OpenWRT_firmware}}. After the installation, OpenWrt provides two commands: \texttt{sendData} and \texttt{recvCSI}. These commands are called from the terminal shown in Fig. \ref{fig:openwrt_ssh}. The basic usages of these commands are given in Fig. \ref{fig:sendData} and Fig. \ref{fig:recvCSI}. The terminal of OpenWrt is achieved through Secure Shell (SSH) from a personal computer (PC) while LuCI can be obtained by any web browser.

The \texttt{sendData} and \texttt{recvCSI} commands are the workhorse of this project. That is why, they deserve a proper explanation:

\begin{itemize}
    \item \texttt{sendData}: This command is responsible to start the routine of sending appropriate pilot symbols to the receiver. As can be seen from Fig. \ref{fig:sendData}, it is required to enter three arguments with the command:
          \begin{enumerate}
              \item \texttt{ifName}: Demands the interface name of the network. This name is designated as \texttt{wlan0} throughout the project.
              \item \texttt{DstMacAddr}: Demands the media access control (MAC) address of the destination (or in our case the receiver). This address is \texttt{C0:4A:00:CB:E1:78} for the receiver router and hardcoded to it, i.e. it is unchangeable.
              \item \texttt{NumOfPacketToSend}: Demands the total number of packets to send to the receiver. After achieving the total number, the routine terminates itself. In order to avoid the interruptions during the measurements due to this termination, the value of this argument is chosen as \texttt{10000000}. The transmission of this amount of packets lasts about one and a half hours which is sufficient for the measurements.
          \end{enumerate}
    \item \texttt{recvCSI}: This command starts the routine of listening to the broadcast of CSI packets assigned to its MAC address. If there is a stream of packets, it saves the log of these packets to the internal memory of the router with the name designated by
          \texttt{output\_file} argument. It continues its routine until an external interrupt from the user (\texttt{Ctrl+C}) comes.

\end{itemize}

The collected logs of the CSI packets with the \texttt{recvCSI} command are transferred to the PC through Ethernet with the following line of code entered to the command prompt:

\begin{verbatim}
    scp <routerName> <routerIP>:<logFileName> <logDir>
\end{verbatim}

This command returns a copy of the CSI packets to the destination designated with \texttt{<logDir>}. However, the copied logs are not in a readable format. Luckily, Atheros CSI Tool developers provide a MATLAB\textsuperscript{\textregistered} function named \texttt{read\_log\_file.m} which extracts the CSI, payload, and packet status data from the logs. This function returns a cell structure consists of successful CSI packets. Every cell forms a struct which contains the following fields:

\begin{itemize}
    \item \texttt{timestamp}: The indicator of when the packet is received. It is expressed in $\mu s$.
    \item \texttt{csi\_len}: CSI data length expressed in bytes.
    \item \texttt{channel}: The center frequency of the channel expressed in MHz.
    \item \texttt{err\_info}: Indicator of whether the packet is successfully received or not.
    \item \texttt{noise\_floor}: Stores the noise floor expressed in dB. Set to 0 in the current version. Needs to be updated.
    \item \texttt{Rate}: Stores the data rate of the received packet. The value is an 8-bit integer number and corresponds to one of the data rates designated by the IEEE 802.11 protocol.
    \item \texttt{bandwidth}: Indicates the channel bandwith. \texttt{0} and \texttt{1} correspond to 20MHz and 40MHz.
    \item \texttt{num\_tones}: Stores the number of OFDM subcarriers used for data transmission.
    \item \texttt{nr}: Stores the number of receiving antenna.
    \item \texttt{nc}: Stores the number of transmitting antenna.
    \item \texttt{rssi}: Stores the RSSI of the combination of all active receiver antennas.
    \item \texttt{rssi1, rssi2, rssi3}: Stores the individual RSSI values of each receiver antenna indexed by 1, 2, and 3.
    \item \texttt{payload\_len}: Stores the payload length of the received packet which is expressed in bytes.
    \item \texttt{csi}: Stores the $\texttt{nr}\times\texttt{nt}\times\texttt{num\_tones}$ sized 3 dimensional CSI matrix. The data is in \texttt{complex int16} format.
    \item \texttt{payload}: Stores the payload vector of size \texttt{payload\_len}. It is formatted in \texttt{uint18}.
\end{itemize}
In this project \texttt{channel}, \texttt{bandwidth}, and \texttt{num\_tones} values are designated as 2462MHz, 20MHz and 56, respectively. Other than that, since both the transmitter and the receiver have three antennas, \texttt{nc} and \texttt{nr} values are equal to 3. However, \texttt{nc} and \texttt{nr} values are important because during the transmission, even though the packet is received successfully, some communication links between antenna pairs may be broken. In order to discard these kinds of measurements, packets with \texttt{nc} and \texttt{nr} values lower than 3 are considered as dropped unsuccessful packets and they are not included in the localization processing.

The most important field of the aforementioned packet struct is the CSI matrix without a doubt. This matrix contains $3\times3\times56=504$ CSI values in complex domain. However, these values are in \texttt{complex int16} format, and interpretation of these values in a physical sense requires a normalization. In this aspect, the magnitude values of every CSI packet can be calculated as

\begin{equation}
    \bar{\Omega}_{ijk} =
    \sqrt{\frac
        {(\Omega \circ \Omega^*)_{ijk}}
        {\sum_j^3 \sum_k^{56}
            \frac
            {(\Omega \circ \Omega^*)_{ijk}}
            {56}
        }}
\end{equation}

where $\Omega$ and $\bar{\Omega}$ represents the raw and normalized CSI matrices, respectively. Here, $\circ$ operation corresponds to the Hadamard matrix product and subindex $ijk$ refers to $i^{th}$, $j^{th}$, and $k^{th}$ elements in the corresponding dimensions of the indexed matrix.

\begin{figure}[!h]
    \centering
    \vspace{5mm}
    \hspace*{-1.2cm}
    \includegraphics[scale = 0.45]{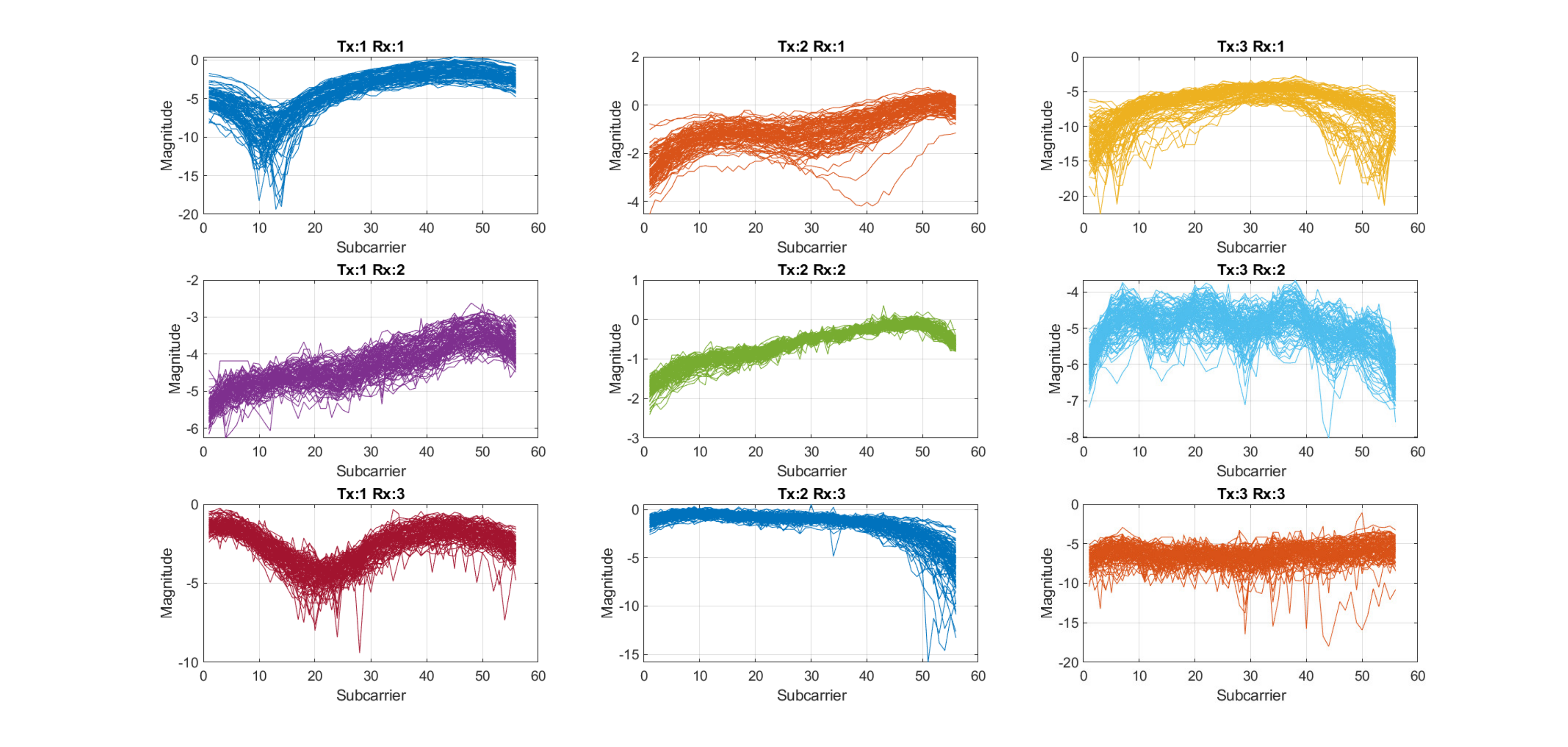}
    \vspace{5mm}
    \caption{Visualization of an example CSI measurement.}
    \label{fig:magPlot}
\end{figure}

\begin{figure}[!h]
    \centering
    \vspace{5mm}
    \hspace*{-1.2cm}
    \includegraphics[scale = 0.45]{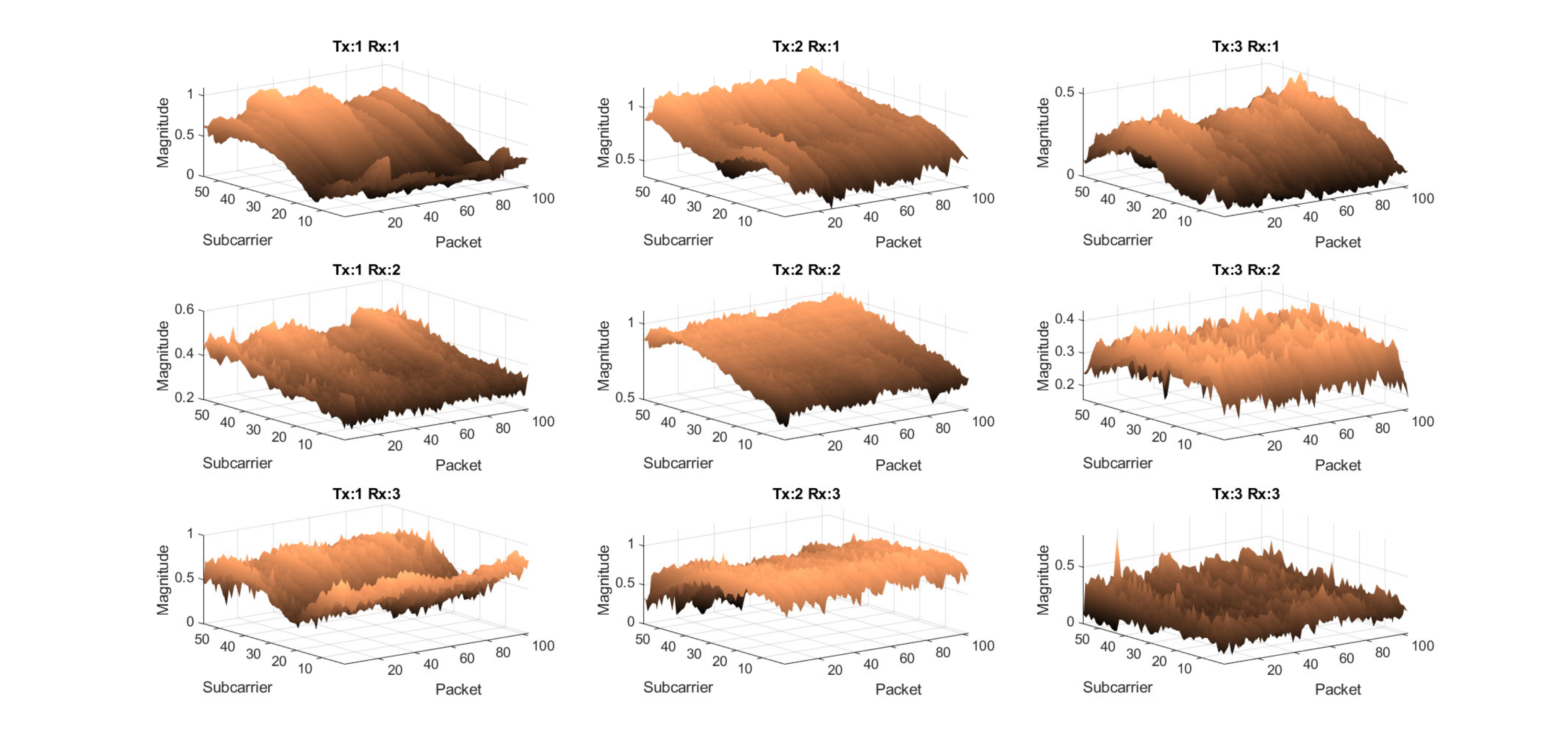}
    \vspace{5mm}
    \caption{Visualization of an example CSI measurement with spectrograms.}
    \label{fig:spect}
\end{figure}

\begin{figure}[!h]
    \centering
    \vspace{5mm}
    \hspace*{-1.2cm}
    \includegraphics[scale = 0.45]{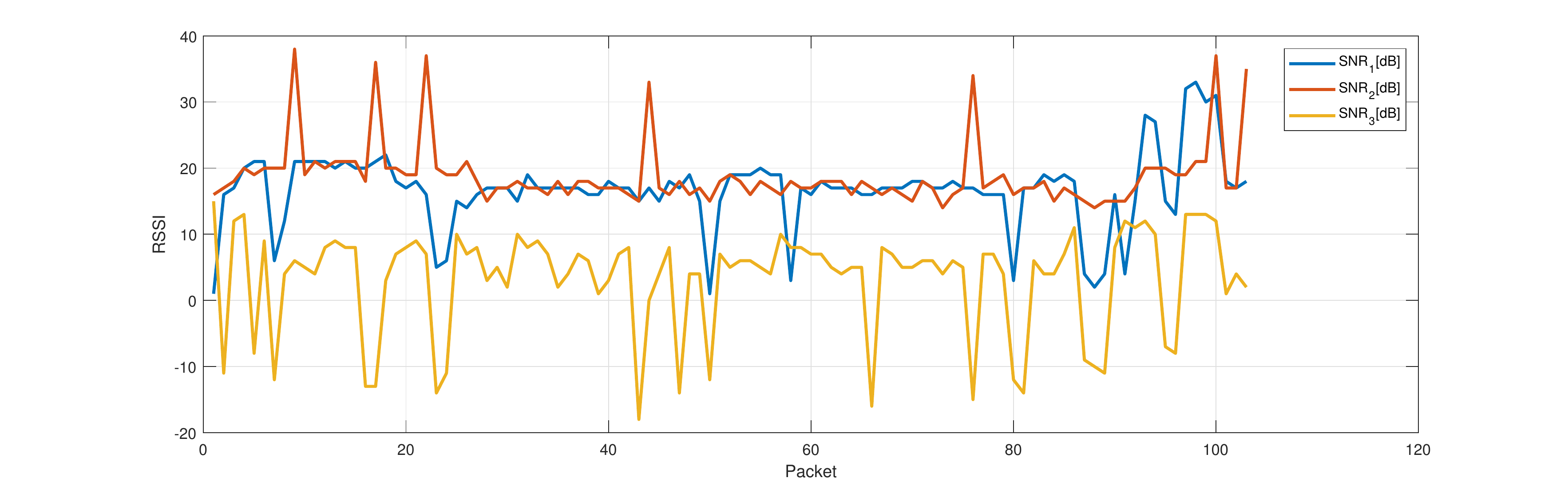}
    \vspace{5mm}
    \caption{Visualization of an example RSSI measurement.}
    \label{fig:rssi}
\end{figure}

An example measurement consists of 60 seconds of broadcasting from the transmitter to the receiver was held and a total number of 103 packets were successfully transmitted. The normalized CSI magnitude data of this example measurement is visualized in Fig. \ref{fig:magPlot}. Here, the CSI data of the successful packet transmissions are plotted logarithmically on top of each other. In order to obtain a better visual inspection, these data are plotted as spectrogram graphs as given in Fig. \ref{fig:spect}. It can be seen that both of these figures reflect the characterization of the current wireless channel. One can conclude this by observing the similarity in the general shapes of every antenna pair link throughout the measurement. On the other hand, one can deduce from the RSSI data given in Fig. \ref{fig:rssi} that RSSI does not contain such rich information as CSI.

\begin{figure}[!h]
    \centering
    \vspace{5mm}
    \includegraphics[scale = 0.55]{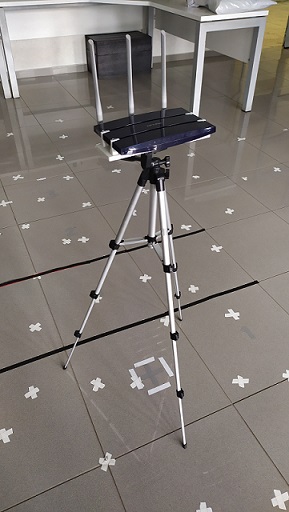}
    \vspace{5mm}
    \caption{The tripod integration of the transmitter.}
    \label{fig:tripod}
\end{figure}

\newpage
Since the required CSI and RSSI data are obtainable, software infrastructure is concluded. Hardware structure, on the other hand, requires a last touch to be concluded. In order to obtain healthy CSI measurements, both of the transmitter and the receiver units should be above from the floor more than a certain degree of height. The reason for this, the floor of the environment acts as a reflective surface and similar scatterings occur for different locations which are highly undesirable. Since the receiver is immobile, a proper placement with an adequate height is sufficient to avoid these unwanted effects. However, taking measurements from different locations requires the mobility of the transmitter. In order to ease the measurements with the mobile transmitter in a constant height, a tripod integration has been done to the transmitter unit as shown in Fig. \ref{fig:tripod}. The attachment of the transmitter to the tripod was done with a custom design apparatus who was printed in 3D.

\section{Localization Data Collection Process}

In the last section, the steps of the functionalization of the CSI data collecting process was given both software- and hardware-wise. After this point, the systematic methodology to collect the CSI data for different locations in the environment has to be specified.

\begin{figure}[!h]
    \centering
    \vspace{5mm}
    \includegraphics[scale = 0.8]{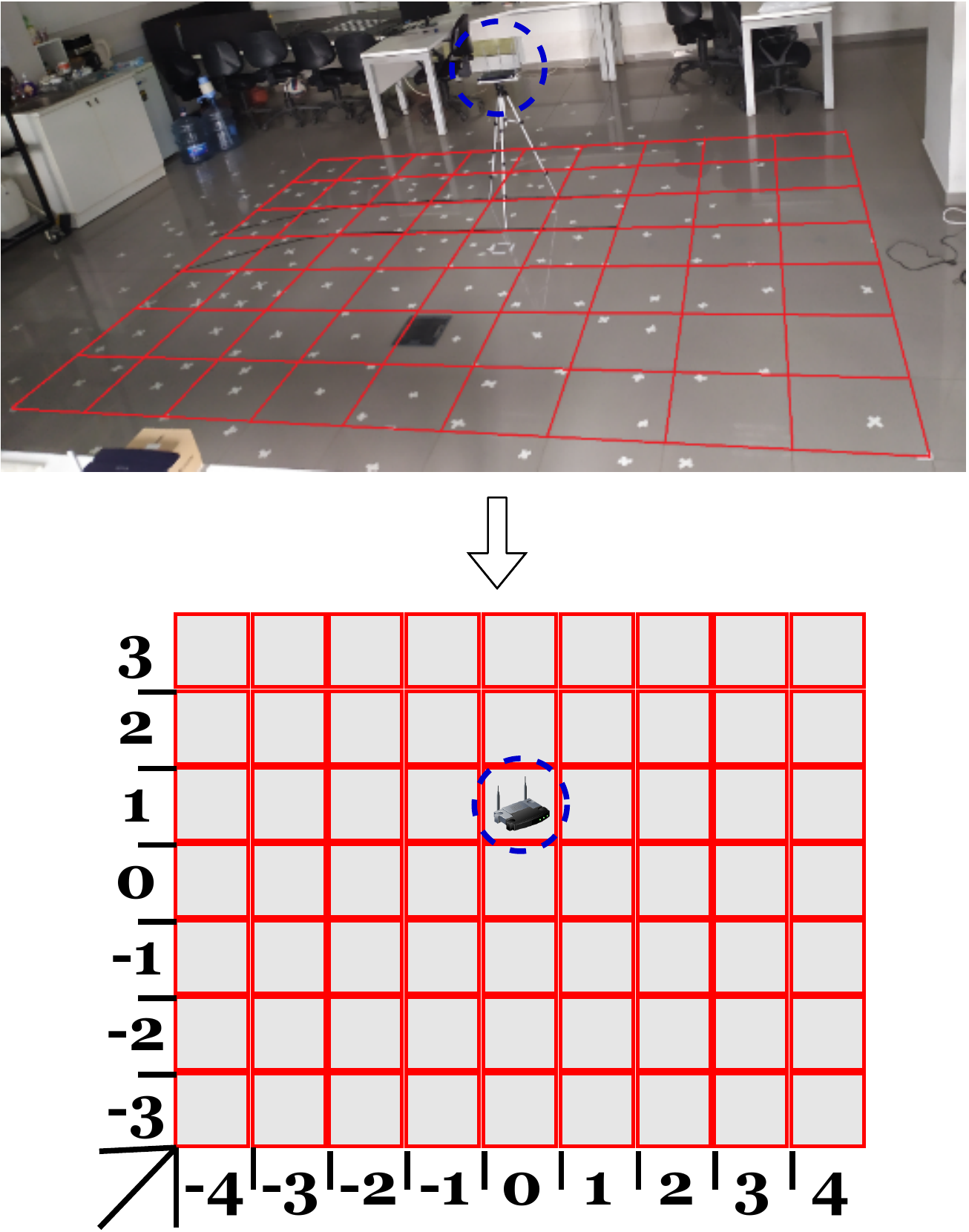}
    \vspace{5mm}
    \caption{Correspondence of transmitter location to the designated coordinates.}
    \label{fig:coords_lab}
\end{figure}

\begin{figure}[!h]
    \centering
    \vspace{5mm}
    \includegraphics[scale = 0.5]{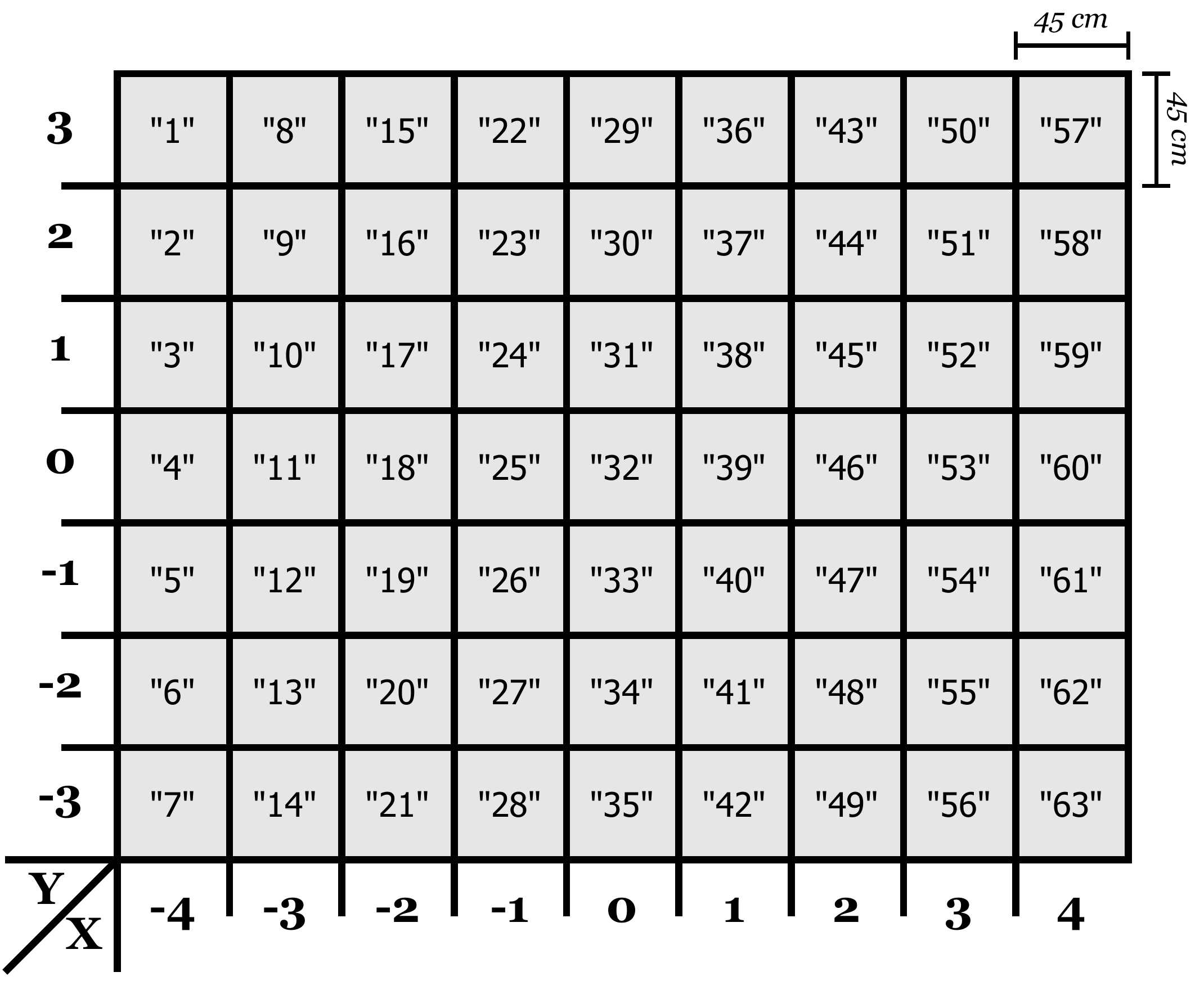}
    \vspace{5mm}
    \caption{Assignment of the coordinates to the environment and the corresponding labels per grid.}
    \label{fig:coords}
\end{figure}

First of all, the environment of interest has to be specified. As mentioned before, all of the experiments were conducted in Istanbul Technical University Artificial Intelligence and Intelligent Systems Laboratory. An area of $4.05m\times3.15 m$ was chosen as the environment of interest. This area divided into a total number of 63 $45cm\times 45 cm$ grids. The mapping from the real environment to the fictional localization coordinates is given in Fig. \ref{fig:coords_lab}. The same coordinate frame with the assigned 63 labels also given in Fig. \ref{fig:coords}. These assignments are crucial for localization processing techniques in the following chapter.

\begin{figure}[!h]
    \centering
    \vspace{20mm}
    \includegraphics[scale = 0.7]{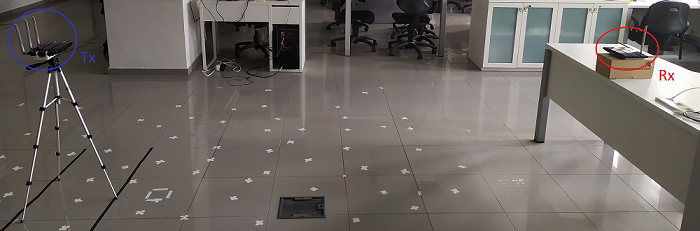}
    \vspace{5mm}
    \caption{Side-view of the experiment setup.}
    \label{fig:txrx}
\end{figure}

\begin{figure}[!h]
    \centering
    \vspace{5mm}
    \includegraphics[scale = 0.7]{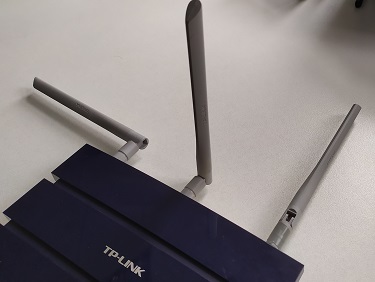}
    \vspace{5mm}
    \caption{Antenna orientations of the receiver.}
    \label{fig:antennas}
\end{figure}

It was mentioned before that the receiver is immobile and it has to be positioned in a constant position with an adequate hight. According to this, the side-view of the whole experiment setup is formed as given in Fig. \ref{fig:txrx}. It can be seen that both of the units have heights respected to the floor close to each other. Another aspect of the receiver is that it is decided to adjust the antenna orientations given in Fig. \ref{fig:antennas}. The main reason for this is the fact that every antenna has a certain electromagnetic polarization and they are designed to work optimally in those polarizations. Therefore, adjusting the receiver antennas in an orthogonal way lowers the correlation between the receiver antennas and this arises a richer CSI data for every location.

After validating the functionality of the setup, the CSI data collection process was initialized. The main goal is to collect a sufficient amount of successful CSI packets from each grid. In this aspect, measurements of one minute were taken from each cell on the first day of the measurements. In order to automize the process, a MATLAB\textsuperscript{\textregistered} script was written. This script accepts the current $(X,Y)$ coordinates of the receiver from the Command Window, and then sends the \texttt{recvCSI} command to the router through SSH. After waiting 60 seconds, it terminates the SSH command and sends \texttt{scp} command to the Command Prompt in order to save the CSI data to PC. Lastly, it runs the \texttt{read\_log\_file} function to check the number of total successful packets and waits for another $(X,Y)$ coordinate input.

\begin{figure}[!ht]
    \centering
    \vspace{5mm}
    \hspace*{-4.5cm}
    \includegraphics[scale = 0.65]{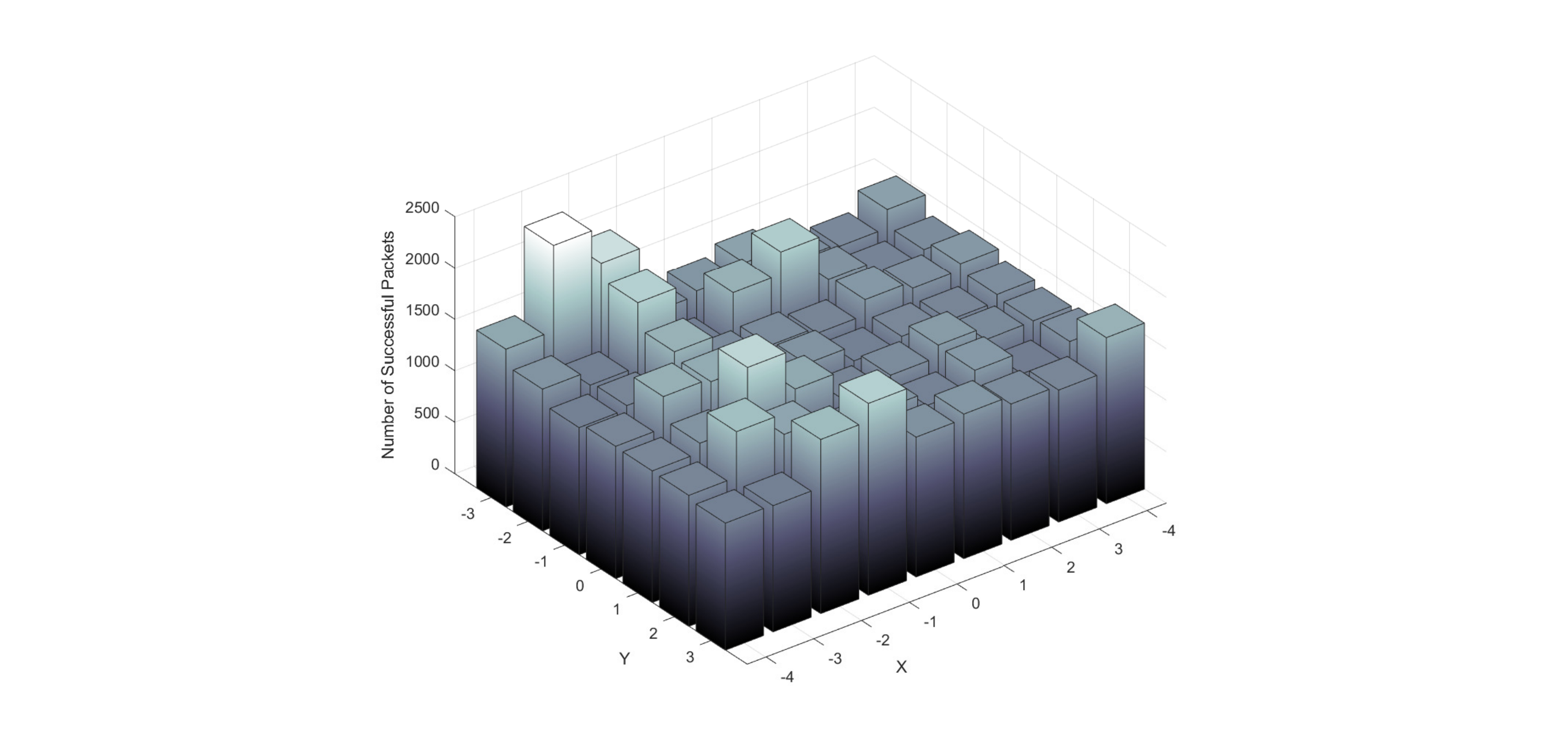}
    \vspace{5mm}
    \caption{Distribution of successful CSI packet measurements with respect to corresponding coordinates.}
    \label{fig:succ}
\end{figure}

The inspection of the first-day measurements showed that there is a heavy imbalance of the number of successful packets between certain areas of the environment. This result shows that these certain areas induce \emph{deep-fade} effects, i.e. destructive combinations of the electromagnetic scatterings from the surrounding objects are more probable. In order to cope with this consequence, additional measurements were taken on four different days. These measurements were taken in a way that the total number of cumulative successful measurements should exceed a predefined limit for that day. At the end of the fourth measurement day, the total number of collected successful packets for each grid is represented as a bar graph in Fig. \ref{fig:succ}. All in all, it is calculated that
\begin{itemize}
    \item the minimum number of successful packets is 1208,
    \item the maximum number of successful packets is 2365,
    \item the median of the number of successful packets is 1343,
    \item and the total number of successful packets is 89677.
\end{itemize}
It is decided that this amount of data is enough to initiate a Deep Learning based modeling which is explained in the following chapter.

\chapter{DATA PROCESS WITH DEEP LEARNING}

In this chapter, the processing of the collected CSI data into position vectors, or grid labels, is discussed. Deep Learning is chosen for the data processing task and the problem is handled both as a regression and a classification problem. In order to achieve better performance, a data pre-processing is conducted. Lastly, the results of the training are visualized and discussed.

\section{Basics of Deep Learning}

In this section, the fundamentals of Deep Learning are given. Additionally, state of the art tools and methods used in this project for improving the performance of Deep Learning are explained. This section is crucial to understand the network architectures given in the next section.

Deep Learning can be thought of as the training of a highly nonlinear parametric model in an enormously high dimensional parameter space. This parameter model is generally chosen as a Deep Neural Network with lots of layers stacked together. A general analytical expression of these networks can be given as

\begin{equation}
    \mathbf{y} = f_L\left(\mathbf{W}^Lf_{L-1}\left(\mathbf{W}^{L-1}\ldots f_1\left(\mathbf{W}^1\mathbf{x}+\mathbf{b}^1\right)\ldots+\mathbf{b}^{L-1}\right)+\mathbf{b}^L\right)
    \label{eqn:fcmodel}
\end{equation}

where $\mathbf{x}$ is the input vector and $\mathbf{y}$ is the output vector of the model. Here, $f_l$, $\mathbf{W}_l$ and $\mathbf{b}_l$ represent the nonlinear activation function, linear multiplication weight matrix and bias vector of the $l^{th}$ layer of the network, respectively. Every operation accomplishes the affine transformation is named as \emph{fully connected layer} (FCL). Therefore, (\ref{eqn:fcmodel}) can be rewritten as

\begin{equation}
    \mathbf{y} = f_L\left(f_{FCN}^L\left(f_{L-1}\left(f_{FCN}^{L-1}\ldots f_1\left(f_{FCN}^1\left(\mathbf{x}\right)\right)\ldots\right)\right)\right)
    \label{eqn:fcn2}
\end{equation}

where

\begin{equation}
    f_{FCN}^l\left(\mathbf{x}^{l}\right) = \mathbf{W}^l\mathbf{x}^{l}+\mathbf{b}^l.
\end{equation}

Here, $\mathbf(x)^{l}$ represents the input of the $l^{th}$ layer. As it can be deduced from (\ref{eqn:fcn2}), a Deep Neural Network consists of a chain of an arbitrary number of consecutive linear and nonlinear operations. Eventually, these Deep Learning schemes aim to optimize the model parameters $\mathbf{W}^{1,\ldots,L}$ and $\mathbf{b}^{1,\ldots,L}$ so that the network accomplishes the desired task with a certain degree of confidence. Following 

\subsection{Deep Learning Tasks}
\label{sect:tasks}

Deep Learning is a data-driven methodology and it is desired that the trained model shows a satisfying performance on a given task with unseen data. Two common data processing tasks for Deep Learning schemes are \emph{regression} and \emph{classification}. The rest of the tasks are out of the scope of this project.

\begin{itemize}
    \item Regression: Mapping of input $\mathbf{x}$ to a continuous output value $y$ as $f: \mathbb{R}^d \rightarrow \mathbb{R}$ is named as regression. Here, $d$ is the dimensionality of the input data. This task can be thought of as fitting a hyper-plane to given data. In terms of Deep Learning, this hyper-plane is highly nonlinear and it has a huge amount of degrees of freedom (learnable parameters). Conversion of the parametric model given in (\ref{eqn:fcmodel}) into a regression model can be done by replacing the output activation function with identity function as $f_L(x)=x$. In other words, the output layers of network architectures for regression tasks are assigned as FCLs.
    
    \item Classification: The mapping $f: \mathbb{R}^d \rightarrow \left\{\pi_1,\pi_2,\ldots,\pi_m\right\}$ is named as classification. Here, the range set $\left\{\pi_1,\pi_2,\ldots,\pi_m\right\}$ contains $m$ different classes and function $f$ chooses one of them according to the input $\mathbf{x}$.  However, in order to fit a mathematical model for $f$, the elements of the range set should be represented as vectors. This representation is held by mapping the set into $m$ dimensional standard basis vectors. Therefore, every element of the vector coincides with a certain class. The implication of this into the Deep Neural Network is done by assigning the output activation function as \emph{softmax} operation, i.e. $f_L(x)=softmax\left(x\right)$. Softmax operation is a nonlinear normalization operation whose output is a vector with elements are in the interval [0 1] and summed up to 1. $i^{th}$ element of the output vector of softmax operation can be calculated as
    \begin{equation}
        y_i = \frac{e^{x^i}}{\sum_j e^{x^j}}.
    \end{equation}
    After obtaining the output vector $\mathbf{y}$, the class prediction can be done as 
    \begin{equation}
        \tilde\pi = \argmax \mathbf{y}.
    \end{equation}
    Thus, the classification task can be thought of as fitting a highly nonlinear and high-dimensional decision plane with lots of degrees of freedom.
\end{itemize}

\subsection{Training of Deep Neural Networks}
\label{sect:train}
Training in Deep Learning terminology corresponds to the optimization of the network parameters $\mathbf{W}^{1,\ldots,L}$ and $\mathbf{b}^{1,\ldots,L}$. The data-driven nature of the methodology requires to form the search space by data. This formation for \emph{supervised} learning tasks, such as regression and classification, defined as the error between the network predictions and the ground-truth data. This error is used to construct a loss function to be minimized with training. 

A common loss function for the regression task is \emph{mean squared error} and defined as

\begin{equation}
    J = \frac{1}{N}\sum_i^N \left(y^{(i)}-\tilde y^{(i)}\right)^2
\end{equation}

where $N$ is the total number of data points, $y^{(i)}$ is the $i^{th}$ ground truth output, and $\tilde y^{(i)}$ is the $i^{th}$ network prediction. For the classification task, on the other hand, one of the most widely used lost function is \emph{cross-entropy loss} and it is calculated as

\begin{equation}
    J = -\sum_i^N y^{(i)}\log\tilde y^{(i)}.
\end{equation}

The minus sign at the beginning is put into there to convert the optimization into a minimization problem.

After setting the appropriate loss function, the learnable parameters of the network are updated iteratively. This update can be done by various optimizer algorithms such as Stochastic Gradient Descent with Momentum, Root Mean Squared Propagation, and Adam. In this project, Adam optimizer is utilized and it is explained in the following sections. For now, the weight update can be given generically as

\begin{equation}
    w^l_{ij}(t+1) = w^l_{ij}(t) +
     f_{optimizer}\left(\frac{\partial J}{\partial w^l_{ij}(t)};\mathbf{\beta}\right) 
\end{equation}

where $w^l_{ij}(t)$ represents the weight in the $i^{th}$ row and the $j^{th}$ column of the weight matrix $\mathbf{W}^l$ at iteration $t$ and $\mathbf{\beta}$ is the pre-defined hyper-parameter vector of the optimizer function $f_{optimizer}$. This update is induced to the network layers by well-known backpropagation algorithm.

\section{Data Pre-processing and Network Architectures}

In this section, the preparation of the CSI data and the proposed Deep Neural Network architectures, both for regression and classification tasks, are given. In order to internalize the architectures in a better way, the state of the art methods and tools are explained, also.

\subsection{Data Preparation}

As it was mentioned in the last chapter, the collected CSI data are in the form of a multi-dimensional matrix with a size of $3\times3\times56$. A common practice for data pre-processing is normalizing the data so that the network is not biased towards the features with higher magnitudes. In other words, it is desired that the network predicts the coordinates, or classes, according to the shape of the CSI.

\begin{figure}[!h]
    \centering
    \vspace{5mm}
    \hspace*{-20mm}
    \includegraphics[scale = 0.5]{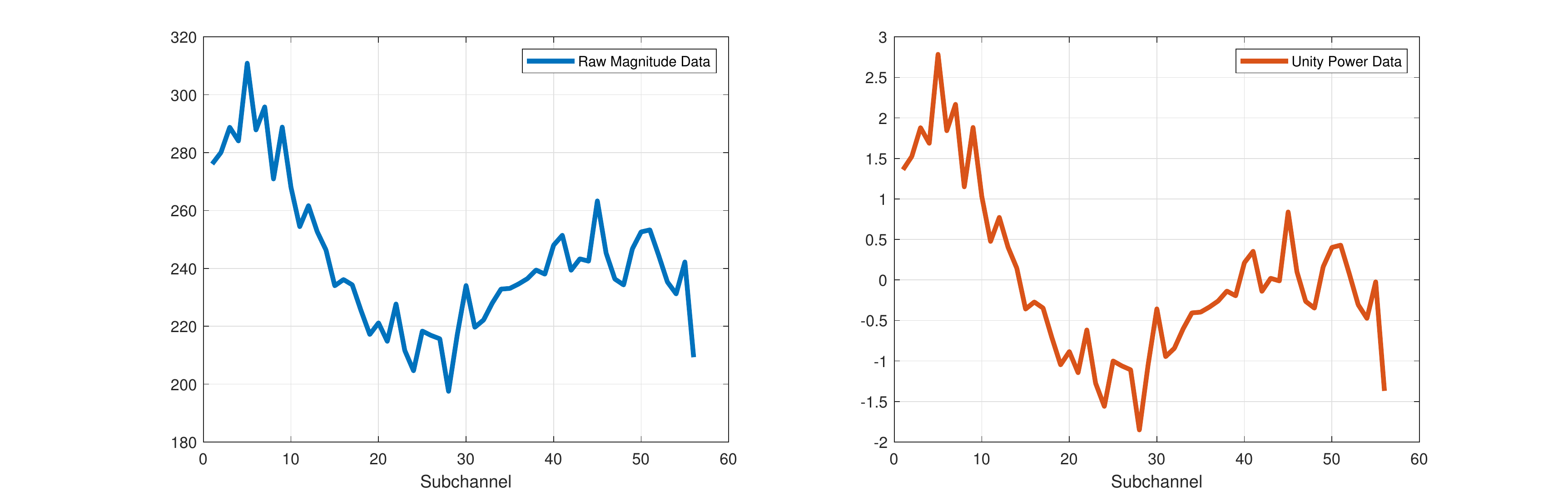}
    \vspace{5mm}
    \caption{Normalizing the data to unit power.}
    \label{fig:powerNorm}
\end{figure}

The normalization of the CSI data is done in a way that every CSI sample between each antenna pair has unit power. This normalization can be done as

\begin{equation}
    \hat{\Omega}_{ij} = \frac{\bar{\Omega}_{ij}-\frac{1}{56}\sum_k^{56}\bar{\Omega}_{ij}}{\sqrt{\frac{1}{56}\sum_k^{56}(\bar{\Omega}_{ij}-\frac{1}{56}\sum_k^{56}\bar{\Omega}_{ij})^2}}
\end{equation}

where $\bar{\Omega}_{ij}$ and $\hat{\Omega}_{ij}$ represents the CSI magnitude vector and the unity power CSI vector between the $i^{th}$ receiver antenna and the $j^{th}$ transmitter antenna, respectively. The application of the power normalization to an arbitrarily selected data sample is given in Fig. \ref{fig:powerNorm}. One should notice the alteration of the magnitude between two data instances. It is worth to mention that the range of the raw magnitude CSI data given in Fig. \ref{fig:powerNorm} is different than the measurements in Fig. \ref{fig:magPlot} because the plots in Fig. \ref{fig:magPlot} are given in logarithmic scale for the sake of better representation.

\begin{figure}[!h]
    \centering
    \vspace{5mm}
    \hspace*{-20mm}
    \includegraphics[scale = 0.5]{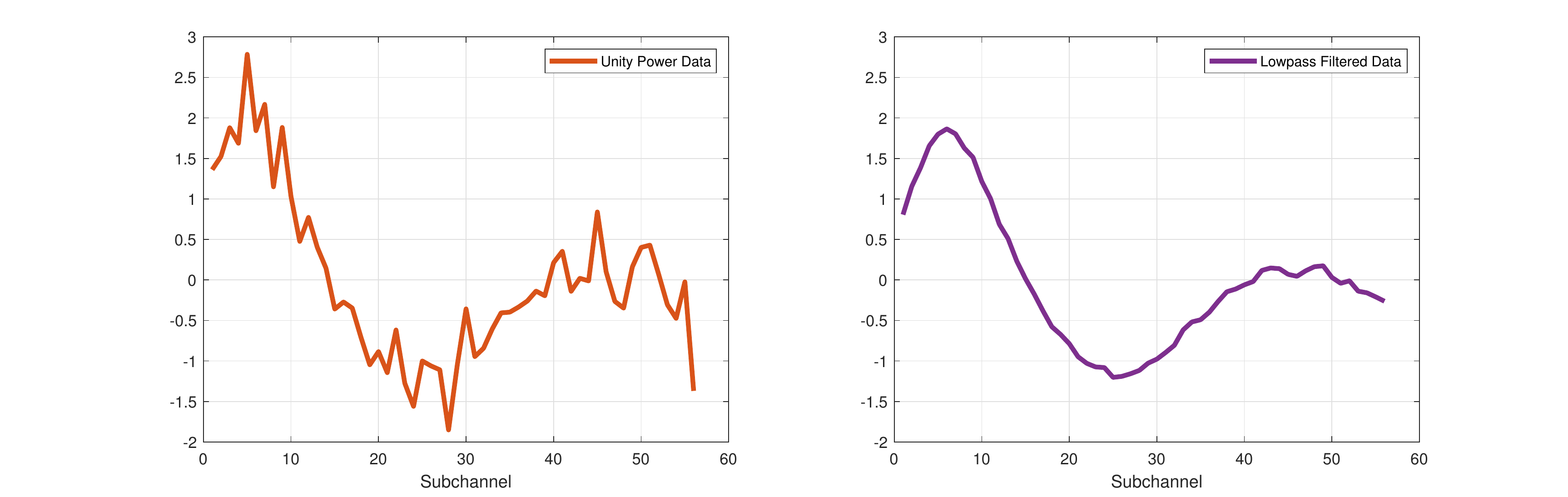}
    \vspace{5mm}
    \caption{Filtering the data with moving average filter.}
    \label{fig:movAvg}
\end{figure}

Another observation can be done with Fig. \ref{fig:powerNorm} is that the CSI magnitude data is relatively noisy. Since it is desired the network to predict by the shape of the CSI data, it is desirable to obtain smooth CSI samples. In order to reduce the noise in data, a moving average filter can be used. The size of the moving window of the filter is chosen as 8, heuristically. The application of this filter into the CSI instance given in \ref{fig:powerNorm} is shown in Fig. \ref{fig:movAvg}. As it can be seen that the lowpass filtering works pretty well to reveal the shape of the CSI data instance.

\begin{figure}[!h]
    \centering
    \vspace{5mm}
    \hspace*{-15mm}
    \includegraphics[scale = 0.5]{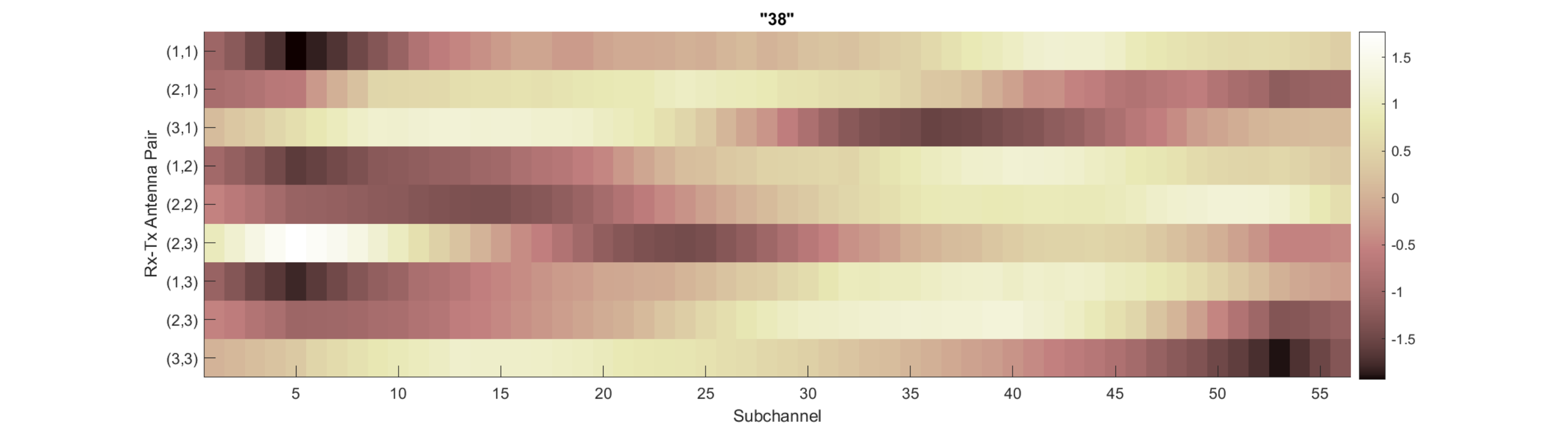}
    \vspace{3mm}
    \caption{Visualization of a pre-processed CSI data instance.}
    \label{fig:figPrint}
\end{figure}

\begin{figure}[!h]
    \centering
    \vspace{5mm}
    \hspace*{-25mm}
    \includegraphics[scale = 0.46]{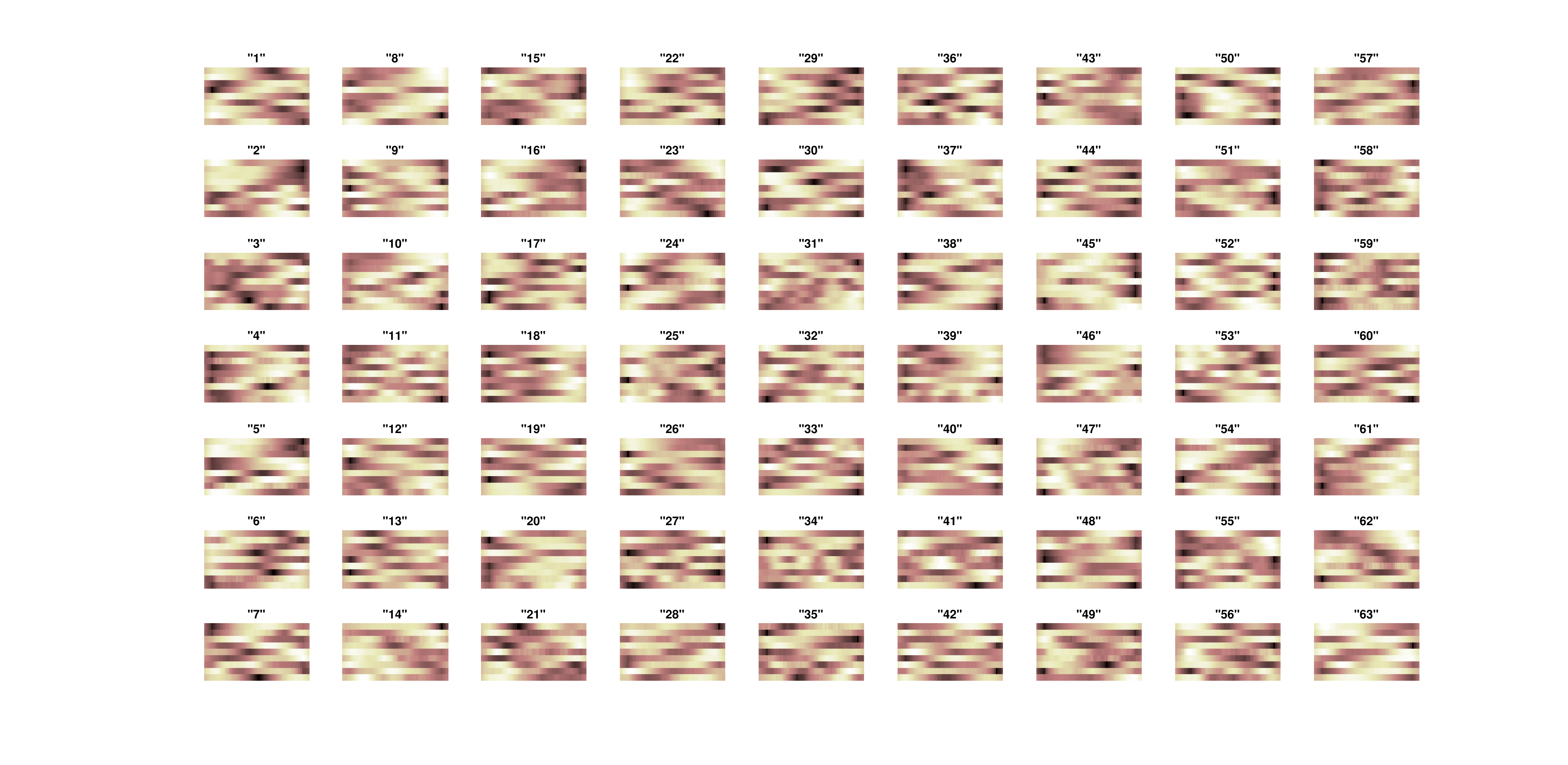}
    \vspace{0mm}
    \caption{Visualization of pre-processed CSI data instances for all of the grids.}
    \label{fig:allGridFigPrint}
\end{figure}

The last part of the data preparation for the training is reshaping the data. This is an important step to represent every grid in the test environment as fingerprint images. This process is simply held by converting the normalized and filtered multidimensional CSI matrices with the sizes of $3\times3\times56$ into 2-dimensional matrices with the sizes of $9\times56$. This operation is visualized in Fig. \ref{fig:figPrint} by taking an instance CSI data of an arbitrary grid ("38") and coloring the magnitude values after reshaping. This image can be thought of as an instance of the probabilistic fingerprint of the grid "38". The application of the same process to the rest of the grids is visualized in Fig. \ref{fig:allGridFigPrint}. As it can be seen that the instances of fingerprints are pretty distinguishable. However, the decision of which fingerprint belongs to which grid is not possible just by inspection. The expectation from Deep Neural Network architecture to capture the underlying relations among these fingerprints. 

\subsection{Deep Neural Network Architecture Designs}

The design of the Deep Neural Networks for this project has been done with the consideration of the common state of the art techniques. These techniques are employed so that the networks perform considerably well. As the first part of this subsection, these techniques and tools are briefly explained. Later, the designed network architectures utilize these methods are investigated.

\subsubsection{Utilized Techniques for The Design}

The first technique to be mentioned is the usage of \emph{convolutional layers} (CLs)\cite{convLeCun}. These layers differ from their FCL counterparts by their ability to capture spatial relationships of the data. Therefore, they are adequate to be used to recognize the fingerprints of the environment.

CLs accomplish this spatial recognition by the employment of a certain number of 2-D linear convolutional filters. This notion resembles with the usage of \emph{masks} in image processing applications. However, in contrast to the image processing applications, the parameters of these filters are included in the training of the neural network rather than being hand-designed. Thus, every element of the filters can be thought as neurons with learnable weights.

\begin{figure}[!h]
    \centering
    \vspace{5mm}
    \includegraphics[scale = 0.35]{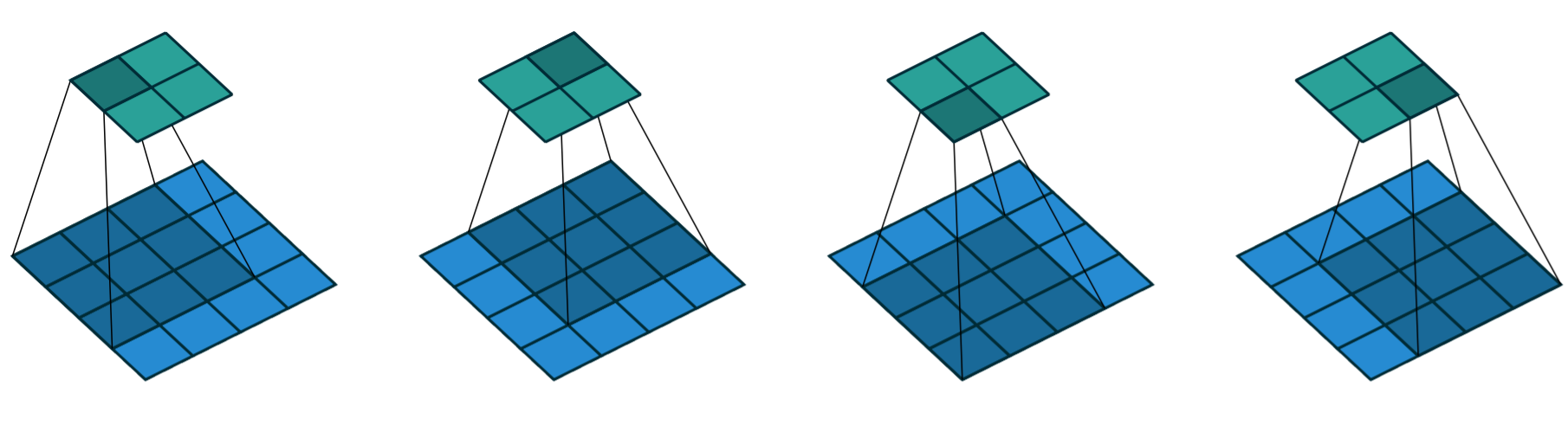}
    \vspace{0mm}
    \caption{Illustration of a $3\times3$ convolutional filter with $(1,1)$ stride [29].}
    \label{fig:convStr1}
\end{figure}

\begin{figure}[!h]
    \centering
    \vspace{5mm}
    \includegraphics[scale = 0.35]{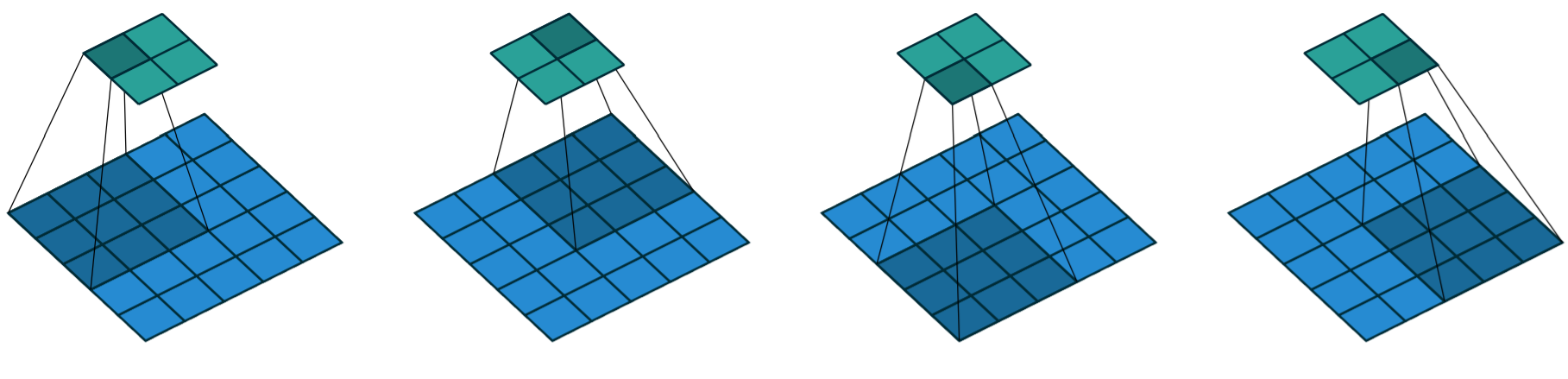}
    \vspace{0mm}
    \caption{Illustration of a $3\times3$ convolutional filter with $(2,2)$ stride [29].}
    \label{fig:convStr2}
\end{figure}

Convolutional filters, or layers, have various design parameters. The two of them are included in the design process of CLs of the proposed networks: \emph{filter size} and \emph{stride}. Filter size is a tuple consists of the height and the width of the filter. These values directly affect the receptive field of the filter. Stride, on the other hand, defines the traverse step sizes of the filter. Since a filter can have different step sizes in different directions, stride can be defined as a tuple consists of horizontal and vertical step sizes. Stride heavily affects the size of the output image so that bigger strides result in smaller output images. These two design parameters are illustrated in Fig. \ref{fig:convStr1} and Fig. \ref{fig:convStr2}. As it can be seen that even though the input image is bigger in Fig. \ref{fig:convStr2}, the output is in equal sizes in both of the figures. This can be justified with the output size formula of the convolutional filter given as\cite{conv}
\begin{equation}
    o = \frac{i-f}{s}+1
\end{equation}
where $o$, $i$, $f$ and $s$ represent the output size, the input size, the filter size, and the stride.

\begin{figure}[!h]
    \centering
    \includegraphics[scale = 0.3]{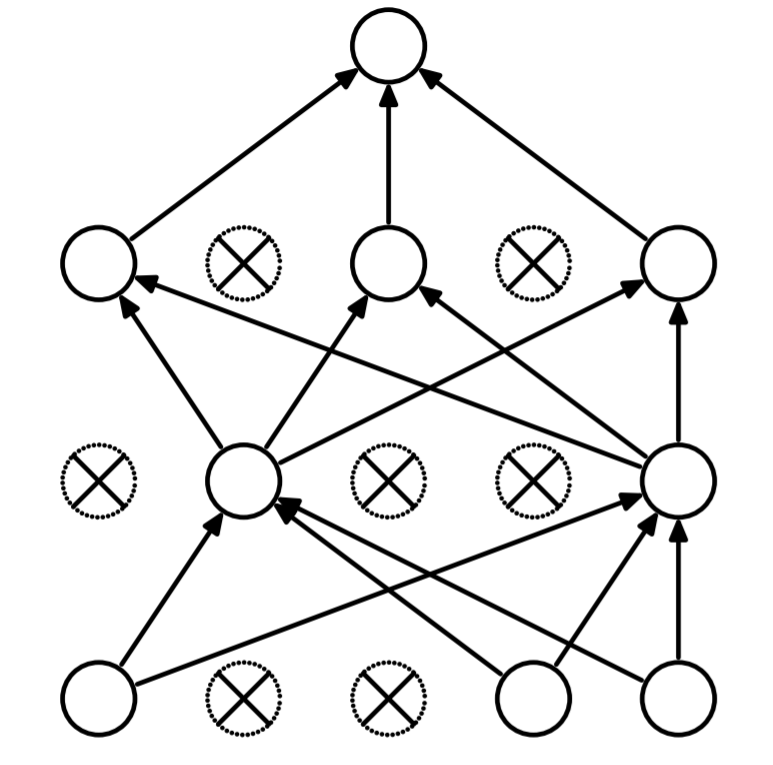}
    \vspace{5mm}
    \caption{Illustration of an instance of dropout operation [30].}
    \label{fig:dropout}
\end{figure}

The second utilized technique in the design of the networks is the usage of \emph{dropout layer}\cite{dropout}. The basic working principle of dropout is the cancellation of the weights in the former layer of the dropout layer during the training time with a predefined probability. In more detail, in every iteration, uniformly distributed random numbers are assigned to weights in the former layers of the dropout layer. The weights with random numbers lower than the predefined probability are canceled out. In this way, the bias of the network towards individual weights is lowered. Thus, the application of dropout reduces the overfitting of the network to the training data. A sample Illustration of the dropout operation for a fully connected network is given in Fig. \ref{fig:dropout}.

\begin{figure}[!h]
    \centering
    \includegraphics[scale = 0.6]{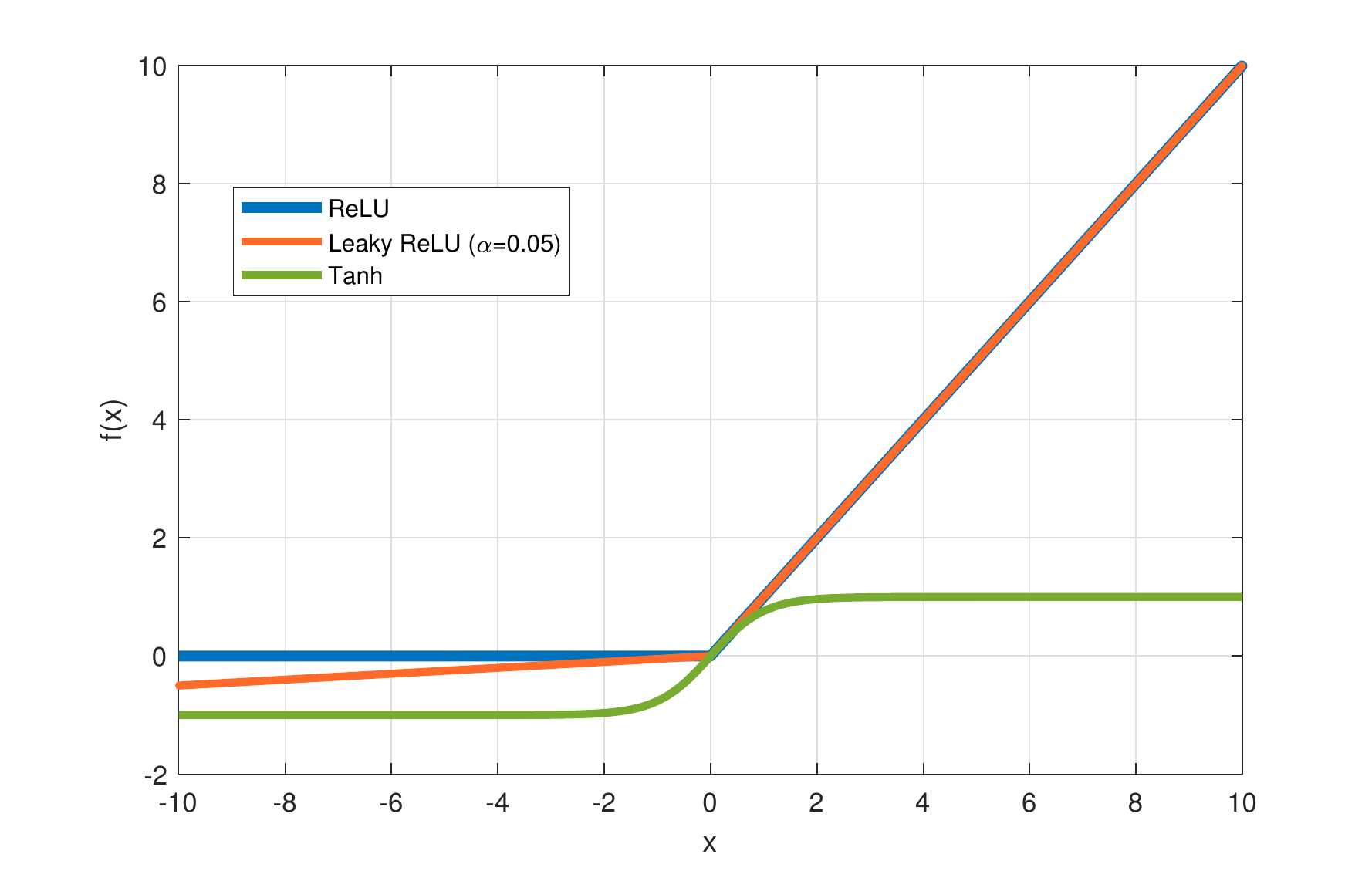}
    \vspace{5mm}
    \caption{Comparison of the activation functions ReLU, Leaky ReLu ($\gamma = 0.05$) and Tanh.}
    \label{fig:lrelucomp}
\end{figure}

The third of the techniques that are utilized in the proposed Deep Neural Networks is the usage of \emph{leaky rectified linear unit} (leaky ReLU) \cite{leakyRelu}. This unit is chosen as the activation function for all of the layers in the proposed networks. The mathematical formulation of this function is given as
\begin{equation}
    f(x) = 
    \begin{cases}
               x, & x \geq 0 \\
        \gamma.x, & x < 0 \\
    \end{cases}
    \label{eqn:leakyrelu}
\end{equation}

where $\gamma$ is a predefined scaling hyper-parameter. The illustration of the leaky ReLU activation function and other common activation functions \emph{Tanh} and \emph{ReLU}\cite{relu} is given in Fig. \ref{fig:lrelucomp}.

The main drawback of the Tanh activation function is that it saturates the inputs with absolute values more than approximately 1.5, considerably. Therefore, the network loses its ability to distinguish the inputs out of this range. Moreover, since the derivatives around these points are so close to zero, the weights are not properly updated. ReLU, on the other hand, solves the saturation problem for positive values and stands out with its computational simplicity by its formulation $f(x) = max(0,x)$. However, cancellation of negative inputs arises similar problems with Tanh. Leaky ReLU solves both of the saturation and the zero gradient problems by introducing a piecewise linear activation function as given in (\ref{eqn:leakyrelu}). Nevertheless, one should be careful about the selection of the hyper-parameter $\gamma$ since the closer it is to 1, the activation function loses its nonlinearity more.

The last technique to be mentioned in this subsection is the optimizer \emph{Adam}\cite{adam}. This optimizer was proposed as the combination of the two other popular optimizers, \emph{Stochastic Gradient Descent with Momentum} (SGDM)\cite{sgdm} and \emph{Root Mean Squared Propagation} (RMSProp)\cite{rmsprop}. It means that Adam uses both of the first and the second moments of the gradients to update the learnable parameters. This can be expressed by the following set of operations:

\begin{equation}
    g(t) = \frac{\partial J(t)}{\partial \theta(t)}
\end{equation}
\begin{equation}
    m(t) = \beta_1.m(t-1)+(1-\beta_1).g(t)
\end{equation}
\begin{equation}
    v(t) = \beta_2.v(t-1)+(1-\beta_2).g^2(t)
\end{equation}
\begin{equation}
    \theta(t) = \theta(t-1) - \eta\frac{m(t)}{\sqrt{v(t)}+\epsilon}
\end{equation}

Here, $g(t)$ represents the gradient of objective function $J(t)$ with respect to the learnable parameter $\theta(t)$ at iteration $t$. $m(t)$ and $v(t)$ correspond to the first and second moment estimates of the gradient, respectively. The hyper-parameters $\beta_1$ and $\beta_2$ are indicators of the exponential decay rates for moment estimation. For example, a first-moment estimation with $\beta_1 = 0.9$ takes approximately the past 10 gradients into consideration. Lastly, $\eta$ and $\epsilon$ values are the conventional learning rate and an arbitrarily small number, such as $10^{-8}$, for numerical consistency. This adaptive nature of Adam optimizer stands out for its contribution to faster and more reliable training processes.

\subsubsection{Network Architectures}

In this project, the localization of the transmitter unit via CSI data is handled both as a regression task and a classification task. In this aspect, the ground-truth output data are formed in two ways. For the regression task, output data is formed as coordinate tuples according to the values given in Fig. \ref{fig:coords}. In order to achieve a more intuitive training monitoring, these tuples are converted into units of meter, e.g. $(3,1)\rightarrow(1.35m,0.45m)$. For the classification task, on the other hand, the labels assigned in Fig. \ref{fig:coords} are used as classes and converted into one-hot vectors, e.g.
\begin{equation}
    \pi_{``63"} \equiv \left[0\quad 0\quad \ldots \quad 0 \quad 1\right]^T.
\end{equation}

\begin{figure}[!h]
    \centering
    \vspace{5mm}
    \hspace*{-15mm}
    \includegraphics[scale = 2]
    {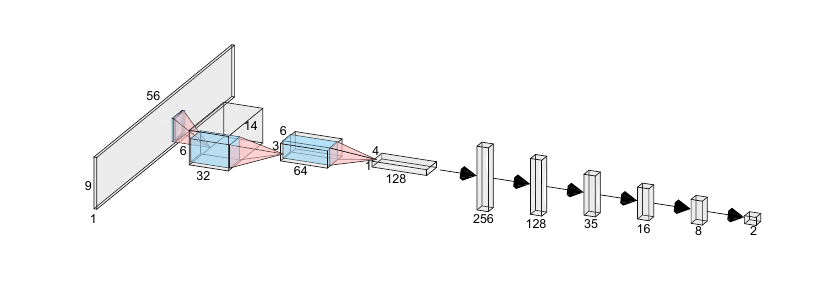}
    \caption{The network architecture for the regression task.}
    \label{fig:regNet}
\end{figure}

\newpage
The proposed network architecture for the regression task is illustrated in Fig. \ref{fig:regNet}. The layer specifications are provided in Table \ref{tbl:layers}(a). The hyper-parameters of leaky ReLU activation functions are chosen as $\gamma=0.01$ for all of the layers. Additionally, the threshold probabilities of cancellation for dropout layers are all set to 0.4.

One should notice that the regression network output is a two-dimensional vector. However, in Subsection \ref{sect:tasks}, the regression task is defined as $f: \mathbb{R}^d \rightarrow \mathbb{R}$. Accordingly, in Subsection \ref{sect:train}, mean squared error loss is defined in terms of scalar outputs. Therefore, with a slight abuse of the definition of regression, the loss function for the proposed regression network defined as 

\begin{equation}
    J = \frac{1}{N}\sum_i^N \left(y^{(i)}-\tilde y^{(i)}\right)^2 + \left(x^{(i)}-\tilde x^{(i)}\right)^2
\end{equation}

where $x^{(i)}$, $y^{(i)}$, $\tilde x^{(i)}$ and $\tilde y^{(i)}$ represent the predicted and the ground-truth vertical and horizontal coordinates of the $i^{th}$ data sample, respectively.

\begin{figure}[!h]
    \centering
    \vspace{1cm}
    \hspace*{-10mm}
    \includegraphics[scale = 1.6]
    {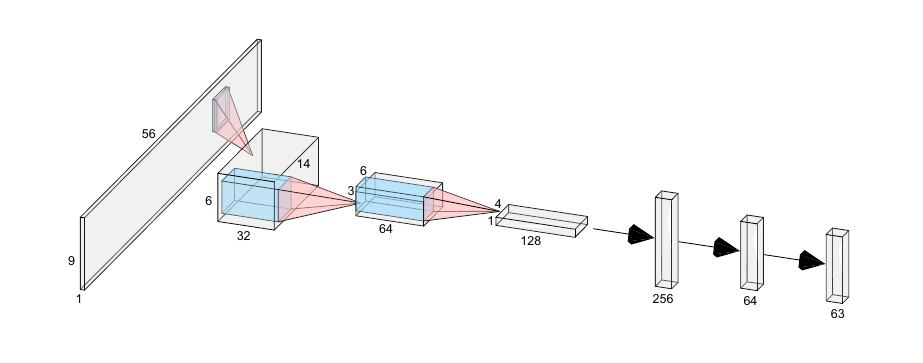}
    \caption{The network architecture for the classification task.}
    \label{fig:classNet}
\end{figure}

The proposed network architecture for the classification task is illustrated in Fig. \ref{fig:classNet} and network layers are specified in Table \ref{tbl:layers}, in a similar manner with the regression network. The same hyper-parameter adjustments with the regression network are followed. For the sake of a better comparison between two networks, the stack of FCLs at the end reduced and adjusted so that the total number of learnable parameters of both of the layers are approximately the same.

\begin{table}[!h]
    \vspace{3cm}
    \caption{The layer specifications of the (a) regression and (b) classification network architectures.}
    \resizebox{\textwidth}{!}{
    \begin{tabular}{ll}
    \hline
    \rowcolor[HTML]{EFEFEF} 
    \multicolumn{1}{|l|}{\cellcolor[HTML]{EFEFEF}\textbf{(a) Regression Network}}      & \multicolumn{1}{l|}{\cellcolor[HTML]{EFEFEF}\textbf{(b) Classification Network}}  \\ \hline
    \multicolumn{1}{|l|}{Input Layer $9\times56\times1$}                               & \multicolumn{1}{l|}{Input Layer $9\times56\times1$}                               \\
    \multicolumn{1}{|l|}{\begin{tabular}[c]{@{}l@{}}CL (32 filters of size $4\times4$ with stride (1,4))\\ Leaky ReLU\\ Dropout\end{tabular}} &
      \multicolumn{1}{l|}{\begin{tabular}[c]{@{}l@{}}CL (32 filters of size $4\times4$ with stride (1,4))\\ Leaky ReLU\\ Dropout\end{tabular}} \\
    \multicolumn{1}{|l|}{\begin{tabular}[c]{@{}l@{}}CL (64 filters of size $4\times4$ with stride (1,2)) \\ Leaky ReLU\\ Dropout\end{tabular}} &
      \multicolumn{1}{l|}{\begin{tabular}[c]{@{}l@{}}CL (64 filters of size $4\times4$ with stride (1,2)) \\ Leaky ReLU\\ Dropout\end{tabular}} \\
    \multicolumn{1}{|l|}{\begin{tabular}[c]{@{}l@{}}CL (128 filters of size $3\times3$ with stride (1,1)) \\ Leaky ReLU \\ Dropout\end{tabular}} &
      \multicolumn{1}{l|}{\begin{tabular}[c]{@{}l@{}}CL (128 filters of size $3\times3$ with stride (1,1)) \\ Leaky ReLU\\ Dropout\end{tabular}} \\
    \multicolumn{1}{|l|}{\begin{tabular}[c]{@{}l@{}}FCL 256\\ Leaky ReLU\end{tabular}} & \multicolumn{1}{l|}{\begin{tabular}[c]{@{}l@{}}FCL 256\\ Leaky ReLU\end{tabular}} \\
    \multicolumn{1}{|l|}{\begin{tabular}[c]{@{}l@{}}FCL 128\\ Leaky ReLU\end{tabular}} & \multicolumn{1}{l|}{\begin{tabular}[c]{@{}l@{}}FCL 64\\ Leaky ReLU\end{tabular}}  \\
    \multicolumn{1}{|l|}{\begin{tabular}[c]{@{}l@{}}FCL 35\\ Leaky ReLU\end{tabular}}  & \multicolumn{1}{l|}{\begin{tabular}[c]{@{}l@{}}FCL 63\\ Softmax\end{tabular}}     \\
    \multicolumn{1}{|l|}{\begin{tabular}[c]{@{}l@{}}FCL 16\\ Leaky ReLU\end{tabular}}  & \multicolumn{1}{l|}{}                                                             \\
    \multicolumn{1}{|l|}{\begin{tabular}[c]{@{}l@{}}FCL 8\\ Leaky ReLU\end{tabular}}   & \multicolumn{1}{l|}{}                                                             \\
    \multicolumn{1}{|l|}{FCL 2}                                                        & \multicolumn{1}{l|}{}                                                             \\ \hline
                                                                                       &                                                                                  
    \end{tabular}
    }
   
    \label{tbl:layers}
\end{table}

\newpage

\section{Training of the Designed Architectures}

As the last step of the project, designed Deep Neural Networks were trained with the pre-processed CSI data. All of the network designs and trainings were done with MATLAB\textsuperscript{\textregistered} 2020a Deep Learning Toolbox\textsuperscript{\texttrademark} on a PC that includes Intel\textsuperscript{\textregistered} Core\textsuperscript{\texttrademark} i7-9750HF 2.6GHz CPU, 16GB RAM and
NVIDIA GeForce GTX 1650 GPU. The training algorithm specifications are listed in Table \ref{tbl:train}. These specifications apply for both the regression and classification networks. 

\begin{table}[!h]
    \centering
    \caption{Training specifications for designed architectures.}
    \begin{tabular}{c}
    \rowcolor[HTML]{EFEFEF} \hline
    Number of Training Data Samples: 67,258 (75\%)  \\ \hline
    Number of Validation Data Samples: 8,968 (10\%) \\ \hline
    \rowcolor[HTML]{EFEFEF} 
    Number of Test Data Samples: 13,451 (15\%)  \\ \hline
    Learning Rate: $\eta=10^{-3}$                \\ \hline
    \rowcolor[HTML]{EFEFEF} 
    Gradient Decay Rate: $\beta_1=0.9$             \\ \hline
    Squared Gradient Decay Rate: $\beta_2=0.999$   \\ \hline
    \rowcolor[HTML]{EFEFEF} 
    Minibatch Size: $N=256$      \\\hline                 
    \end{tabular}
    \label{tbl:train}
\end{table}

\begin{figure}[!h]
    \centering
    \vspace{1cm}
    \hspace*{-15mm}
    \includegraphics[scale = 0.78]{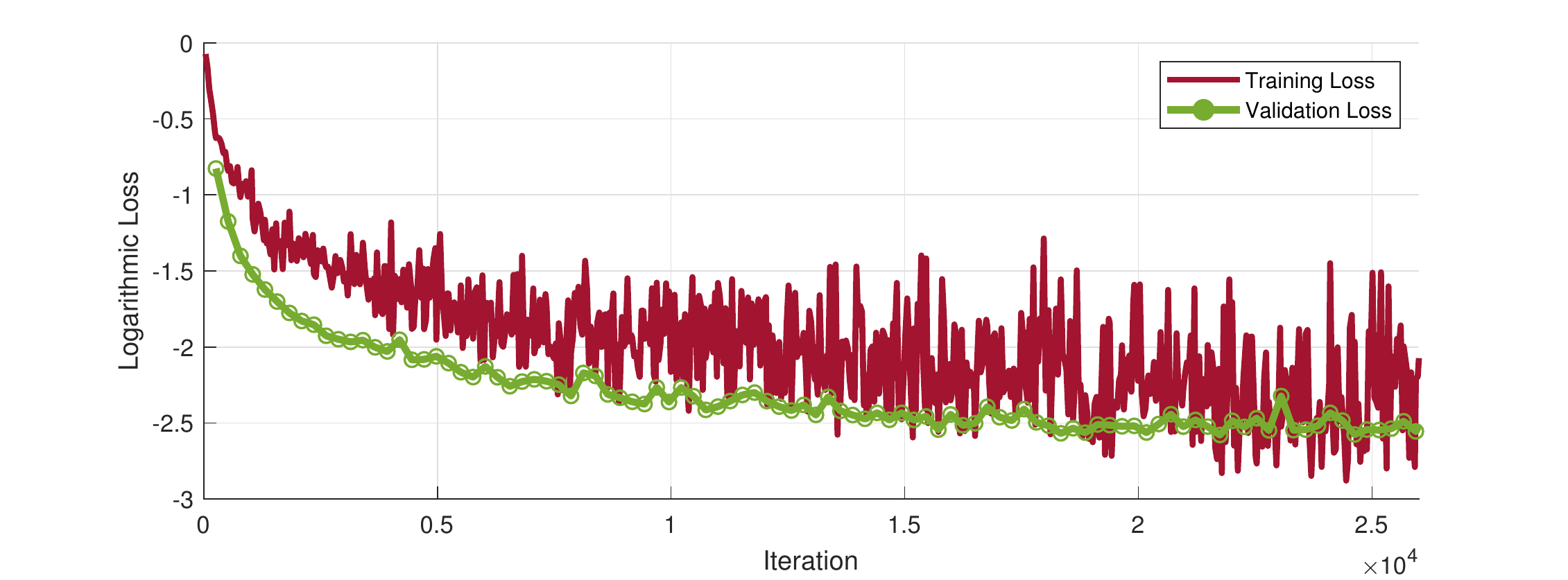}
    \caption{Progression of the logarithmic loss for the regression network.}
    \label{fig:regLoss}
\end{figure}

The progression of the logarithmic loss values through the iterations for the regression network is given in Fig. \ref{fig:regLoss}. The training is held for 100 epochs for this network. One may argue that the training loss has higher values than the validation loss for almost every epoch and there is a problem about the training. However, it should not be forgotten that dropout layers are utilized in the network architecture. At every iteration, some portion of the weights is canceled and this affects the training loss. However, validation loss is calculated with the full utilization of the weights. This effect of dropout explains also the heavy oscillations in the logarithmic training loss towards the end of the training. This is senseful since more the parameters converge to optimum, the network becomes more sensitive to the absence of parameters. However, one should remind that these loss values are in logarithmic scale and these oscillations remain meaningless for linear scale.

\begin{figure}[!h]
    \centering
    \vspace{1cm}
    \hspace*{-1mm}
    \includegraphics[scale = 0.85]{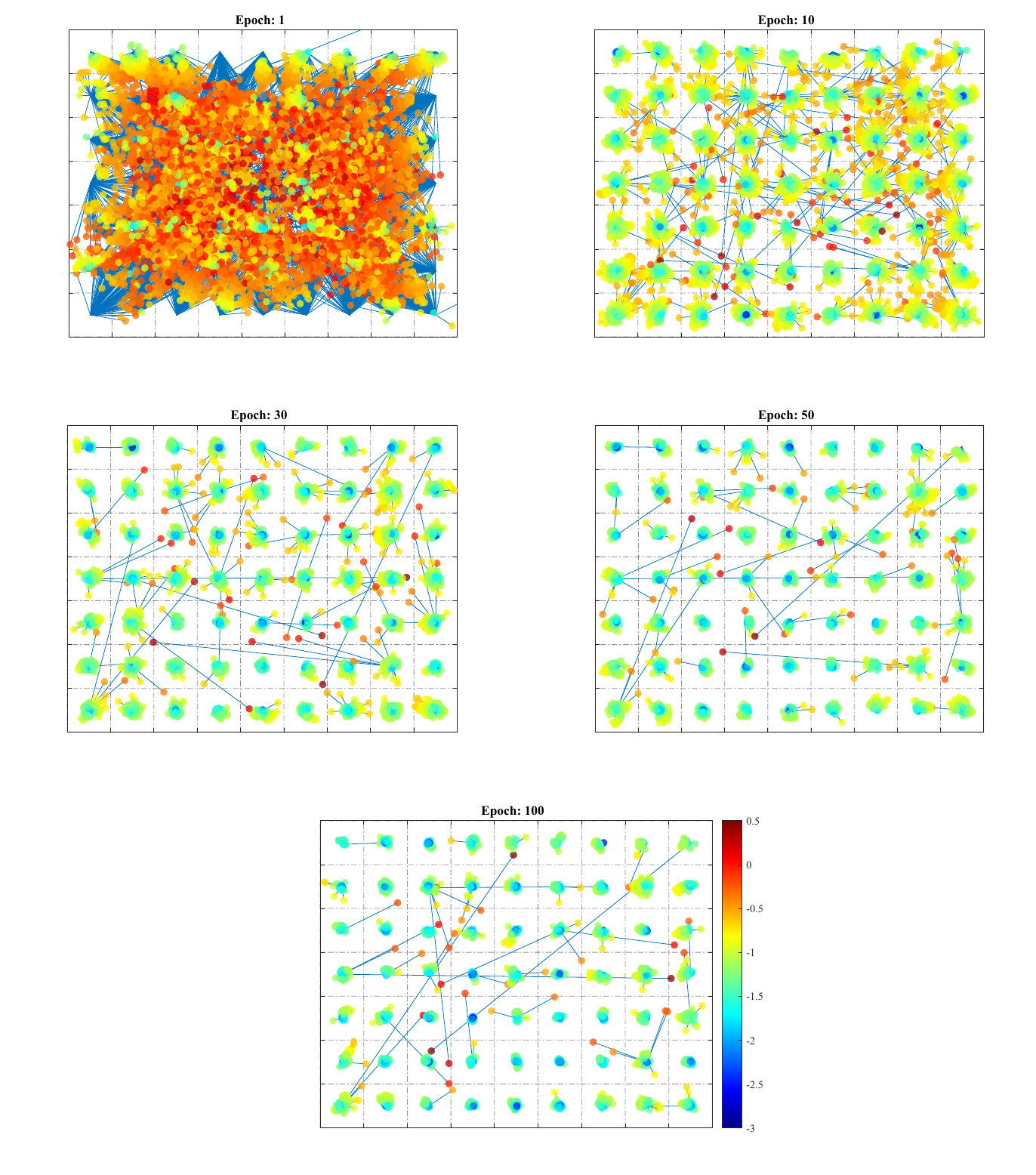}
    \caption{The change of prediction errors through epochs.}
    \label{fig:epochs}
\end{figure}

The prediction errors during the training are visualized in Fig. \ref{fig:epochs}. The colors indicate the logarithmic euclidian distance between the predictions and the ground-truth position data points. As it can be seen that towards the end of the training, only the outlier data stands out among all of the predictions. One may argue that the network stoped learning after 50 epochs. However, the real effect of the training shows itself on the convergence to the ground-truth data, not the compensation of the outliers.

\begin{figure}[!h]
    \centering
    \vspace{1cm}
    \hspace*{0mm}
    \includegraphics[scale = 0.6]{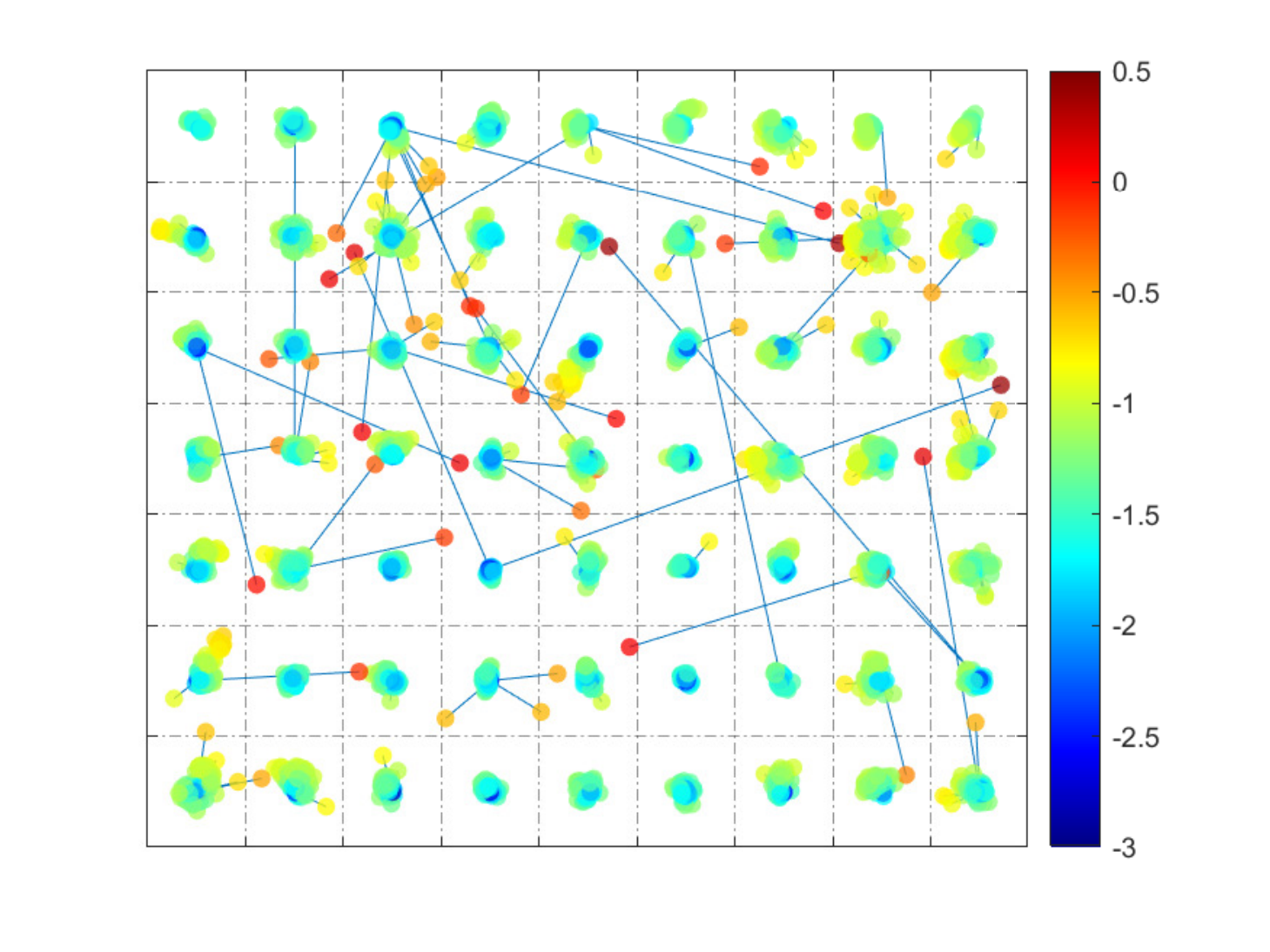}
    \caption{Prediction errors for test data after training.}
    \label{fig:testErr}
\end{figure}

The same visualization applied to the test data and the result is given in Fig. \ref{fig:testErr}. As it can be seen, except the outliers, the network performs well for test data also. This can be seen from the total mean square testing error of 0.027. This value is also calculated for training data as 0.014 and for validation data as 0.040.

\begin{figure}[!h]
    \centering
    \vspace{1cm}
    \hspace*{-15mm}
    \includegraphics[scale = 1.33]{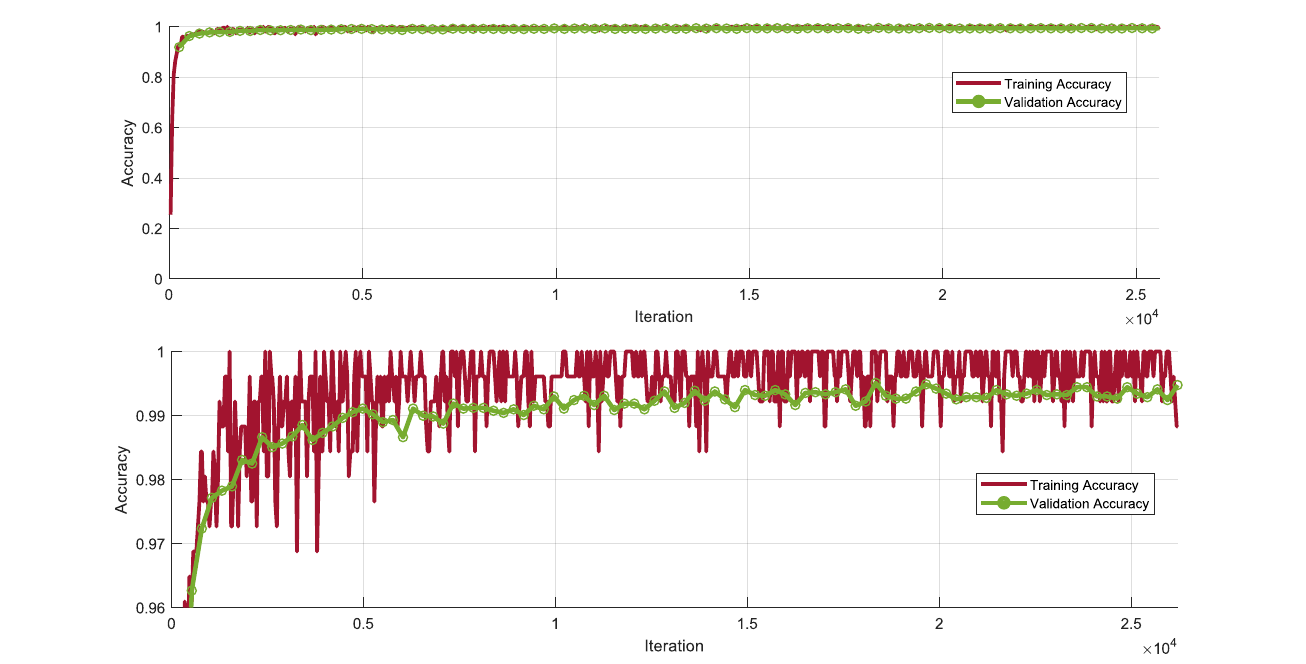}
    \caption{Progression of the accuracy for the classification network.}
    \label{fig:classLoss}
\end{figure}

The investigation of the training of the classification network is done in a similar way with the regression network. Instead of visualizing the cross-entropy loss values, the accuracy values through epochs and their closed up view are given in Fig. \ref{fig:classLoss}. The training is held for 100 epochs just like the training of the regression network. The final accuracy values for the training, validation and test data sets are calculated as 0.9977, 0.9948 and 0.9937, respectively. These accuracies correspond to total numbers of 155, 54 and 85 misclassifications for corresponding data sets, respectively. 

By the inspection of Fig. \ref{fig:epochs} and Fig. \ref{fig:testErr}, it can be said that the regression network slightly outperforms the classification network. This is due to the fact that the prediction error visuals are given in grids and the predictions that exceed their grids are lower than the aforementioned total number of misclassifications. Furthermore, offers a confidence check to the user since the user is aware that the data are collected from the center of the grids. Thus, a deviation from the multiplies of 0.45m in the predicted coordinates result in less confident predictions for the user.
\chapter{CONCLUSION AND FUTURE WORK}

In this project, a framework for CSI based localization has been constructed. CSI is a notion that indicates the current wireless channel effects. In this manner, first, an extensive research has been conducted about basic communication concepts. Wireless channel models and physical layer effects were investigated. Additionally, in order to internalize the CSI notion, the OFDM technique was investigated. After setting the theoretical background, the hardware and software infrastructure was deployed for the CSI data collection task. Two wireless routers with adequate chipsets were utilized in this matter and they were assigned as transmitter and receiver units. Parallelly, the test environment was set and partitioned into equispaced grids. Accordingly, a data collection routine has been conducted and this process is automated with code scripts. After acquiring enough data from each grid, these data are stored for the localization processing task. Deep Learning was chosen as the processing method for localization due to the complex nature of the wireless channels. In this matter, the basics of Deep Learning were researched along with the state of the art methods. After deciding the methods to use, collected CSI data were pre-processed, accordingly. Then, two network architectures were designed so that each of them serves a different Deep Learning task, regression and classification. In other words, the localization task was considered both as a regression problem and a classification problem. After designing the networks properly, the results of these two networks were compared. It was seen that the regression network slightly works better than the classification network.

As the conclusion of this project, it can be said that the usage of the CSI data for localization tasks, especially for indoor applications, is a promising area. The complex nature of the CSI data can be handled with the power of Deep Learning. Moreover, Deep Learning itself is an evolving area and it becomes easier to design and acquire, day by day. Thus, the integration of Deep Neural Networks to the localization systems seems to arise in the near future.

The freshness of the CSI based localization topic opens lots of doors for possible future work. By the influence of this project, some possible research topics can be given. Firstly, all of the experiments in this project were done offline. In order to see the real performance of the localization performance, the testing phase should be conducted in online, also. Secondly, it has been seen that outlier data cause abnormal predictions. During the real-time applications on mobile units, such as UAVs, CSI data should be included in a sensor fusion algorithm. Lastly, other promising Deep Learning methods and architectures in the literature can be employed in order to achieve better generalization and accuracy.

\bibliographystyle{itubib}
\bibliography{Others/Thesis.bib}



\ozgecmis{\vspace{10mm}

\newsavebox{\mysquare}
\savebox{\mysquare}{\textcolor{black}{\rule[2.3pt]{3.4pt}{3.4pt}}}

\setlength{\TPHorizModule}{10pt}
\setlength{\TPVertModule}{10pt}
\begin{textblock}{1}(40,10)
 \begin{figure}[p]
 \includegraphics[scale=0.3,keepaspectratio=true]{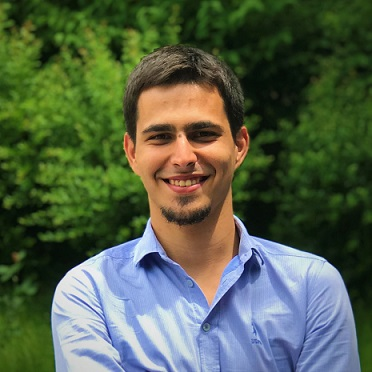}
\end{figure}

\end{textblock}
\textbf{Name Surname:}  Kutay Bölat \\

\vspace{-3mm}
\textbf{Place and Date of Birth:}  Istanbul/Turkey 05.04.1996 \\

\vspace{-3mm}
\textbf{E-Mail:} bolatk@itu.edu.tr\\

\textbf{EDUCATION:} 
\vspace{-3mm}
\begin{itemize}
  \item \textbf{B.Sc.:} 2019, Istanbul Technical University, Faculty of Electrical and Electronics Engineering, Electronics and Communication Engineering Department
  \item \textbf{B.Sc.:} 2020, Istanbul Technical University, Faculty of Electrical and Electronics Engineering, Control and Automation Engineering Department
\end{itemize}

\textbf{PROFESSIONAL EXPERIENCE AND REWARDS:}   
\vspace{-3mm}
\begin{itemize}
  \item 2019- Researcher at Istanbul Technical University Artificial Intelligence and Intelligent Systems (AI2S) Laboratory.
\end{itemize}
}

\end{document}